\documentclass[11pt]{article}
\usepackage{graphicx,verbatim,array,multicol,palatino,algorithm}
\usepackage{amsmath,amssymb,amsfonts,ifthen,setspace,color,
            natbib,epsfig,epstopdf,relsize}
\newboolean{ShowFigures}
\newboolean{UnBlinded}
\newboolean{DoubleSpaced}
\setboolean{ShowFigures}{true}
\setboolean{UnBlinded}{false}
\setboolean{DoubleSpaced}{false}
\def\bzero{\boldsymbol{0}}
\def\bone{\boldsymbol{1}}
\usepackage{amsmath,amssymb,amsfonts}
\def\ba{\boldsymbol{a}}

\def\bc{\boldsymbol{c}}

\def\bg{\boldsymbol{g}}

\def\bu{\boldsymbol{u}}
\def\bv{\boldsymbol{v}}
\def\bw{\boldsymbol{w}}
\def\bx{\boldsymbol{x}}
\def\by{\boldsymbol{y}}

\def\bA{\boldsymbol{A}}
\def\bB{\boldsymbol{B}}
\def\bC{\boldsymbol{C}}

\def\bI{\boldsymbol{I}}

\def\bM{\boldsymbol{M}}

\def\bT{\boldsymbol{T}}

\def\bX{\boldsymbol{X}}

\def\bZ{\boldsymbol{Z}}
\def\thick#1{\hbox{\rlap{$#1$}\kern0.25pt\rlap{$#1$}\kern0.25pt$#1$}}

\def\bbeta{\boldsymbol{\beta}}

\def\btheta{\boldsymbol{\theta}}

\def\bmu{\boldsymbol{\mu}}

\def\bsigma{\boldsymbol{\sigma}}

\def\bSigma{\boldsymbol{\Sigma}}

\def\thick#1{\hbox{\rlap{$#1$}\kern0.25pt\rlap{$#1$}\kern0.25pt$#1$}}
\def\smbalpha{{\thick{\scriptstyle{\alpha}}}}

\def\smbalpha{\widehat{\smbalpha}}

\def\hbar{{\overline h}}

\def\Hsc{{\mathcal H}}

\let\proglang=\textsf
\def\myand{\&\ }

\def\smhalf{{\textstyle{\frac{1}{2}}}}

\def\Diff{{\sf D}}

\def\simind{\stackrel{{\tiny \mbox{ind.}}}{\sim}}
\def\diag{\mbox{diag}}

\def\diagonal{\mbox{diagonal}}

\def\tr{\mbox{tr}}

\def\vecof{\mbox{vec}}

\def\beq{\begin{equation}}
\def\eeq{\end{equation}}
\def\bev{\begin{verbatim}}
\def\eev{\end{verbatim}}

\def\jump{\vskip3mm\noindent}
\def\lboxit#1{\vbox{\hrule\hbox{\vrule\kern6pt
      \vbox{\kern6pt#1\kern6pt}\kern6pt\vrule}\hrule}}

\def\thickboxit#1{\vbox{{\hrule height 1mm}\hbox{{\vrule width 1mm}\kern6pt
          \vbox{\kern6pt#1\kern6pt}\kern6pt{\vrule width 1mm}}
               {\hrule height 1mm}}}
\def\colthickboxit#1{\vbox{{\textcolor{brown}{\hrule height 3mm}}
          \hbox{{\textcolor{brown}{\vrule width 3mm}}\kern6pt
          \vbox{\kern6pt#1\kern6pt}\kern6pt{\textcolor{brown}{\vrule width 3mm}}}
               {\hrule height 5mm}}}

\def\beq{\begin{eqnarray}}
\def\eeq{\end{eqnarray}}
\def\beqn{\begin{eqnarray*}}
\def\eeqn{\end{eqnarray*}}

\def\xnew{x_{\mbox{{\tiny new}}}}
\def\ynew{y_{\mbox{{\tiny new}}}}

\def\bse{\begin{eqnarray*}}
\def\ese{\end{eqnarray*}}
\def\raybe{\begin{eqnarray}}
\def\rayee{\end{eqnarray}}

\def\fat#1{\hbox{\rlap{$#1$}\kern0.25pt\rlap{$#1$}\kern0.25pt$#1$}}

\def\punder{\underline{p}}

\def\bse{\begin{eqnarray*}}
\def\ese{\end{eqnarray*}}
\def\raybe{\begin{eqnarray}}
\def\rayee{\end{eqnarray}}

\def\pe+{p_{{\scriptscriptstyle{\rm E}+}}}
\def\pg+{p_{{\scriptscriptstyle{\rm G}+}}}

\def\exerDBtags#1#2{\null}
{\begin{list}
{\bulletcolour$\bullet$}
{\setlength{\leftmargin}{30mm}\setlength{\itemsep}{0.5ex}}\sf}
{\end{list}\normalsize}
{\begin{list}{\small\bulletcolour$\bullet$}
{\setlength{\leftmargin}{0.75in}\setlength{\itemsep}{0.5ex}}\small}
{\end{list}\normalsize}
\definecolor{DarkGreen}{rgb}{0.412,0.545,0.133}
\def\updateMult{F_{{\tiny\mbox{update}}}}

\def\bA{\boldsymbol{A}}
\def\bC{\boldsymbol{C}}
\def\bc{\boldsymbol{c}}

\def\bB{\boldsymbol{B}}
\def\btheta{\boldsymbol{\theta}}

\def\diag{\mbox{diag}}

\def\diagonal{\mbox{diagonal}}
\def\bsigma{\boldsymbol{\sigma}}

\def\simind{\stackrel{{\tiny \mbox{ind.}}}{\sim}}
\def\xnew{x_{\mbox{{\tiny\mbox{new}}}}}
\def\ynew{y_{\mbox{{\tiny\mbox{new}}}}}
\def\bmu{\boldsymbol{\mu}}
\def\bSigma{\boldsymbol{\Sigma}}
\def\bZ{\boldsymbol{Z}}
\def\bbeta{\boldsymbol{\beta}}
\def\bX{\boldsymbol{X}}
\def\bu{\boldsymbol{u}}
\def\bv{\boldsymbol{v}}
\def\bw{\boldsymbol{w}}
\def\bx{\boldsymbol{x}}

\def\by{\boldsymbol{y}}
\def\bI{\boldsymbol{I}}
\def\bzero{\boldsymbol{0}}
\def\bone{\boldsymbol{1}}
\def\punder{\underline{p}}
\def\smhalf{{\textstyle{\frac{1}{2}}}}
\def\thickboxit#1{\vbox{{\hrule height 1mm}\hbox{{\vrule width 1mm}\kern6pt
          \vbox{\kern6pt#1\kern6pt}\kern6pt{\vrule width 1mm}}
               {\hrule height 1mm}}}
\def\myand{\&\ }
\def\jump{\vskip3mm\noindent}

\def\nwarm{n_{\mbox{\scriptsize warm}}}

\def\bywarm{\by_{\mbox{\scriptsize warm}}}

\def\bCwarm{\bC_{\mbox{\scriptsize warm}}}

\def\ynew{y_{\mbox{\tiny\mbox{new}}}}

\def\bcnew{\bc_{\mbox{\tiny\mbox{new}}}}

\def\Hsc{\mathcal{H}}

\begin{document}
\ifthenelse{\boolean{DoubleSpaced}}
{\setstretch{1.5}}{}

\begin{center}
{\LARGE\bf Variational inference for count response semiparametric regression}

\vskip4mm
{\sc By J. Luts and M.P. Wand}
\vskip4mm
{\it
School of Mathematical Sciences,
University of Technology Sydney, Broadway 2007, Australia}

\end{center}

\vskip3mm
\centerline{17th September, 2013}
\vskip3mm
\centerline{\sc Summary}
\vskip2mm

Fast variational approximate algorithms are developed for 
Bayesian semiparametric regression when the response
variable is a count, i.e. a non-negative integer.
We treat both the Poisson and Negative Binomial families 
as models for the response variable. Our approach utilizes
recently developed methodology known as non-conjugate
variational message passing. For concreteness, we focus 
on generalized additive mixed models, although our 
variational approximation approach extends to a wide
class of semiparametric regression models such as 
those containing interactions and elaborate 
random effect structure.
\noindent
\jump

\section{Introduction}\label{sec:intro}

A pervasive theme impacting Statistics in the mid-2010s
is the increasing prevalence of data that are big in terms
of volume and/or velocity. One of many relevant articles
is \citet{Michalak12}, where the need for systems that perform 
real-time streaming data analyses is described.
The analysis of high volume data and velocity data requires
approaches that put a premium on speed, possibly at the cost
of accuracy. Within this context, we develop methodology for fast,
and possibly online, semiparametric regression analyses
in the case of count response data.

Semiparametric regression, as defined in \citet{Ruppert09}, is
a fusion between parametric and nonparametic regression that
integrates low-rank penalized splines and wavelets, mixed
models and Bayesian inference methodology.
In \citet{Luts13} we developed semiparametric regression 
algorithms for high volume and
velocity data using a mean field variational Bayes (MFVB)
approach. It was argued there that MFVB, or similar methodology,
is necessary for fast batch and online semiparametric
regression analyses, and that more traditional methods such 
as Markov chain Monte Carlo (MCMC) are not feasible. 
However, the methodology of \citet{Luts13} was restricted
to fitting Gaussian and Bernoulli response models. 
Extension to various other response distributions,
such as $t$, Skew Normal and Generalized Extreme Value
is relatively straightforward using approaches described
in Wand \textit{et al.} (2011). However count response distributions 
such as the Poisson and Negative Binomial distribution have received 
little attention in the MFVB literature. Recently
\citet{Tan13} used an extension of MFVB, known as non-conjugate 
variational message passing, to handle Poisson mixed models
for longitudinal data and their lead is followed here for
more general classes of count response semiparametric
regression models.

In generalized response regression, the Poisson distribution is 
often bracketed with the Bernoulli distribution since both
are members of the one-parameter exponential family. However,
variational approximations for Poisson response models
are not as forthcoming as those with Bernoulli responses.
\citet{Jaakkola00} derived a lower bound on the
Bayesian logistic regression marginal likelihood 
that leads to tractable
approximate variational inference. 
As explained in \citet{Girolami06} and
\citet{Consonni07}, the \citet{Albert93} auxiliary
variable representation of Bayesian probit regression leads
to a different type of variational approximation method for
binary response regression. There do not appear to be 
analogues of these approaches for Bayesian Poisson regression
and different routes are needed. An effective solution is
afforded by a recent extension of MFVB, 
due to \citet{Knowles11}, known as \emph{non-conjugate 
variational message passing}.
The Negative Binomial distribution can also be handled using
non-conjugate variational message passing, via its well-known 
representation as a Poisson-Gamma mixture \citep[e.g.][]{Lawless87}.
We adopt such an approach here and develop MFVB algorithms for
both Poisson and Negative Binomial semiparametric regression models.
For ease of presentation, we restrict attention to the 
special case of generalized additive mixed models, but
extension to other semiparametric regression models is straightforward.

Section \ref{sec:background} lays down required notation
and distributional results. It also provides a brief 
synopsis of non-conjugate
mean field variational Bayes. The models are then described 
in Section \ref{sec:modelDescription}. The article's centerpiece
is Section \ref{sec:VBinference}, which is where the variational
inference algorithms for count response semiparametric regression
are presented. In Section \ref{sec:real-time} we describe real-time 
fitting of such models. Numerical illustrations are given in 
Section \ref{sec:numerical} and an appendix contains derivations
of the aforementioned variational algorithms.

\section{Background Material}\label{sec:background}

The specification of the models and their fitting via
variational algorithms requires several definitions and results,
and are provided in this section.

\subsection{Distributional Definitions}

Table \ref{tab:distribs} lists all distributions used in this
article. In particular, the parametrization of the 
corresponding density functions and probability functions 
is provided.

\begin{table}[ht]
\begin{center}
\begin{tabular}{lll}
\hline
distribution     & density/probability function in $x$  & abbreviation \\[0.1ex]
\hline\\[-0.9ex]
Poisson            &$\lambda^x\,e^{-\lambda}/x!;\quad x=0,1,\ldots$   
&  $\mbox{Poisson}(\lambda)$      \\[2ex]
Negative Binomial            &
$\displaystyle{\frac{\kappa^{\kappa}\Gamma(x+\kappa)\mu^x}
{\Gamma(\kappa)(\kappa+\mu)\Gamma(x+1)}};\  x=0,1\ldots;$   
&  $\mbox{Negative-Binomial}(\mu,\kappa)$      \\[2ex]
&  $\kappa,\mu>0$ & \\[2ex]
Uniform   & $1/(b-a);\quad a<x<b$
& $\mbox{Uniform}(a,b)$ \\[2ex]
Multivariate Normal   & $|2\pi\bSigma|^{-1/2}
\,\exp\{-\smhalf(\bx-\bmu)^T\bSigma^{-1}(\bx-\bmu)\}$
& $N(\bmu,\bSigma)$ \\[2ex]
Gamma            &$\displaystyle{\frac{B^{A}\,x^{A-1}e^{-B\,x}}
{\Gamma(A)}};\quad  x>0;\  A,B>0$   
&  $\mbox{Gamma}(A,B)$      \\[2ex]
Inverse-Gamma            &$\displaystyle{\frac{B^{A}\,x^{-A-1}e^{-B/x}}
{\Gamma(A)}};\quad  x>0;\  A,B>0$   
&  $\mbox{Inverse-Gamma}(A,B)$      \\[2ex]
Half-Cauchy&$\displaystyle{\frac{2\sigma}{\pi(x^2+\sigma^2)}};\quad  
x>0;\  \sigma>0$ & $\mbox{Half-Cauchy}(\sigma)$ \\[2ex]
\hline
\end{tabular}
\caption{\textit{Distributions used in this article and their
corresponding density/probability functions.}}
\label{tab:distribs}
\end{center}
\end{table}

\subsection{Distributional Results}\label{sec:distResults}

The variational inference algorithms given in 
Section \ref{sec:VBinference} make use of the following distributional
results:

\jump
\noindent
{\bf Result 1.} {\em Let $x$ and $a$ be random variables such that}
$$x|\,a\sim\mbox{Poisson}(a)\quad\mbox{\em and}\quad
a\sim\mbox{Gamma}(\kappa,\kappa/\mu).$$
{\em Then $x\sim\mbox{Negative-Binomial}(\mu,\kappa)$.}

\jump
\noindent
{\bf Result 2.} {\em Let $x$ and $a$ be random variables such that}
$$x|\,a\sim\mbox{Inverse-Gamma}(1/2,1/a)\quad\mbox{\em and}\quad
a\sim\mbox{Inverse-Gamma}(\smhalf,1/A^2).$$
{\em Then $\sqrt{x}\sim\mbox{Half-Cauchy}(A)$.}

Result 1 is a relatively well-known distribution theoretic
result \citep[e.g.][]{Lawless87}.
Result 2 is related to established results
concerning the $F$ distribution family, and this particular 
version is taken from \citet{Wand11}.

\subsection{Non-conjugate Variational Message Passing}

Non-conjugate variational message passing \citep{Knowles11}
is an extension of MFVB. It can yield tractable variational
approximate inference in situations where ordinary MFVB 
is intractable.

MFVB relies on approximating the joint posterior density
function $p(\btheta|\by)$ by a product form $q(\btheta) =
\prod_{i=1}^{d} q(\btheta_i)$, where $\btheta$ corresponds to the
hidden nodes in Figure \ref{fig:dag}. The optimal $q$-density
functions, denoted by $q^*(\btheta_i)$,
are those that minimize the Kullback-Leibler divergence
$$
\int q(\btheta) \log\left(\frac{q(\btheta)} {p(\btheta|\by)}\right)\,d\btheta.\\ 
$$
\noindent 
An equivalent
optimization problem represents maximizing the lower bound on the
marginal likelihood $p(\by)$:
\begin{equation}
\punder(\by;q)\equiv\exp \left\{ \int q(\btheta)
\log\left(\frac{p(\btheta,\by)} {q(\btheta)}\right)\,d\btheta\right\}.\\ \nonumber
\end{equation}
The optimal $q$-density functions can be shown to satisfy
\begin{equation}
q^*(\btheta_i) \propto \exp \left[ E_{-\btheta_i}
\left\{ \log p(\btheta_i|\mbox{rest}) \right\} \right], \quad 
1\leq i \leq d, \nonumber \\
\end{equation}
where $E_{-\btheta_i}$ denotes expectation with respect to the density
$\prod_{j\neq i} q_{j}(\btheta_j)$
and `rest' denotes all random variables in the model
other than $\btheta_i$.

In the event that one of the $E_{-\btheta_i}
\left\{\log p(\btheta_i|\mbox{rest})\right\}$ 
is not tractable, let's say the one corresponding to
$q(\btheta_j)$ for some $j\in\{1,\ldots,d\}$,
non-conjugate variational message passing offers a
way out \citep{Knowles11}. It first postulates that $q(\btheta_j)$ is an
exponential family density function with natural parameter vector
$\boldsymbol{\eta}_j$ and natural statistic $\bT(\btheta_j)$.
The optimal parameters are then obtained via updates of
the form
\begin{equation}
\boldsymbol{\eta}_j \leftarrow \left\{ \text{var} \left(\bT(\btheta_j)
\right) 
\right\}^{-1} \left\{ \Diff_{\boldsymbol{\eta_j}}  
E_{\btheta}\left[ \log p(\btheta,\by) \right] \right\},\\
\label{eq:NCVMPupdates}
\end{equation}
where $\Diff_{\bx}f$ is the derivative vector of $f$ with respect to
$\bx$ and $\text{var}(\bv)$ denotes the covariance matrix of random
vector $\bv$ \citep{Magnus99}. \citet{Wand13} derived fully simplified
expressions for (\ref{eq:NCVMPupdates}) in case $q(\btheta_j)$ has a
Multivariate Normal density with mean $\bmu_{q(\theta_j)}$ and covariance matrix
$\bSigma_{q(\theta_j)}$
\begin{equation}
\begin{array}{l}
\bSigma_{q(\theta_j)} \leftarrow \left\{-2 \, \vecof^{-1}
\left(\left[\Diff_{\vecof(\bSigma)} 
E_{\btheta}\left\{\log p(\btheta,\by)\right\}\right]^T \right) 
\right\}^{-1},\\[3ex] 
\bmu_{q(\theta_j)} \leftarrow \bmu_{q(\theta_j)} 
+ \bSigma_{q(\theta_j)} \left[ \Diff_{\bmu} 
E_{\btheta}\left\{\log p(\btheta,\by)\right\} \right]^T \\
\label{eq:updatesMatt}
\end{array}
\end{equation}
with $\text{vec}(\bA)$ denoting a vector formed by stacking the
columns of matrix $\bA$ underneath each other in order from left to
right and $\text{vec}^{-1}(\ba)$ a matrix formed from listing the
entries of vector $\ba$ in a column-wise fashion in order from left to
right.

\section{Model descriptions}\label{sec:modelDescription}

Count responses are most commonly modelled according to the
Poisson and Negative Binomial distributions. The latter may
be viewed as an extension of the former through the introduction
of an additional parameter.

Throughout this section we use $\simind$ to denote 
``independently distributed as''.

\subsection{Poisson additive mixed model}

We work with the following class of Bayesian Poisson additive mixed models:
\begin{equation}
\begin{array}{c}
y_i|\,\bbeta,\bu \simind
\mbox{Poisson}[\exp\{(\bX\bbeta+\bZ\bu)_i\}],\quad 1 \leq i \leq n,\\[2ex]
\bu|\,\sigma_{1}^2,
\ldots,\sigma_{r}^2\sim N(\bzero,\mbox{blockdiag}
(\sigma_{1}^2\,\bI_{K_1},\ldots,\sigma_{r}^2\,\bI_{K_r})),\\[2ex]
\bbeta\sim N(\bzero,\sigma_{\beta}^2\bI_p),\quad\mbox{and}\quad
\sigma_{\ell}\simind\mbox{Half-Cauchy}(A_{\ell}),\quad 1\le\ell\le r.
\end{array}
\label{eq:poissonModel}
\end{equation}
Here $\by$ is an $n\times1$ vector of response variables, 
$\bbeta$ is a $p\times1$ vector of fixed effects, 
$\bu$ is a vector of random effects, $\bX$ and $\bZ$ 
corresponding design matrices, and $\sigma_{1}^2,\ldots,\sigma_{r}^2$
are variance parameters corresponding to sub-blocks
of $\bu$ of size $K_1,\ldots,K_r$.

Result 2 of Section \ref{sec:distResults} allows us to replace
$\sigma_{\ell}\simind\mbox{Half-Cauchy}(A_{\ell})$ with
$$\quad
\sigma_{\ell}^2\,|\,a_{\ell}\simind\mbox{Inverse-Gamma}(\smhalf,1/a_{\ell}),
\quad a_{\ell}\simind\mbox{Inverse-Gamma}(\smhalf,1/A_{\ell}^2),
\quad 1\le\ell\le r,
$$
which is more amenable to variational inference. 

Note that the $r=1$ version of (\ref{eq:poissonModel}) is treated
in \citet{Wand13}.

\subsection{Negative Binomial additive mixed model}

The Negative Binomial distribution is an extension of the 
Poisson distribution in that the former approaches 
a version of the latter as the shape parameter $\kappa\to\infty$ 
(see Table \ref{tab:distribs}). The Negative Binomial
shape parameter allows for a wider range of dependencies
of the variance on the mean and can better handle
over-dispersed count data.

The Bayesian Negative Binomial additive mixed model treated here is
\begin{equation}
\begin{array}{c}
y_i|\,\bbeta,\bu \simind
\mbox{Negative-Binomial}[\,\exp\{(\bX\bbeta+\bZ\bu)_i\}],\quad 1 \leq i \leq n,\\[2ex]
\bu|\,\sigma_{1}^2,
\ldots,\sigma_{r}^2\sim N(\bzero,\mbox{blockdiag}
(\sigma_{1}^2\,\bI_{K_1},\ldots,\sigma_{r}^2\,\bI_{K_r})),
\quad
\bbeta\sim N(\bzero,\sigma_{\beta}^2\bI_p),
\\[3ex]
\quad
\sigma_{\ell}\simind\mbox{Half-Cauchy}(A_{\ell}),\quad 1\le\ell\le r, \
\mbox{and}\quad
\kappa \sim \mbox{Uniform}\left(\kappa_{\text{min}},\kappa_{\text{max}}\right).
\end{array}
\label{eq:negBinModel}
\end{equation}
Courtesy of Result 1 given in Section \ref{sec:distResults},
$$y_i|\,\bbeta,\bu,\kappa \simind
\mbox{Negative-Binomial}[\,\exp\{(\bX\bbeta+\bZ\bu)_i\},\kappa],\quad 1\le
i\le n,$$
can be replaced by
$$
y_i|g_i \simind \mbox{Poisson}\left(g_i \right), \quad 
g_i|\bbeta,\bu,\kappa \simind 
\mbox{Gamma}\left(\kappa,\kappa\,\exp\{-(\bX\bbeta+\bZ\bu)_i\} \right),
\quad 1\le i\le n,
$$
where $\bg$ is the $n\times1$ vector containing the $g_i$, 
$1\le i\le n$.

\subsection{Directed Acyclic Graph Representations}

Figure \ref{fig:dag} provides a directed acyclic graph
representation of models (\ref{eq:poissonModel}) and  (\ref{eq:negBinModel}).
Observed data are indicated by the shaded node while parameters, 
random effects and auxiliary variables are so-called hidden nodes. This visual
representation shows that the Poisson case and Negative
Binomial case have part of their graphs in common. The locality
property of MFVB \citep[e.g. Section 2 of][]{Wand11} means
that the variational inference algorithms for the two models 
have some components in common. We take advantage of this
in Section \ref{sec:VBinference}.

\ifthenelse{\boolean{ShowFigures}}
{
\begin{figure}[!ht]
\centering
{\includegraphics[width=0.65\textwidth]{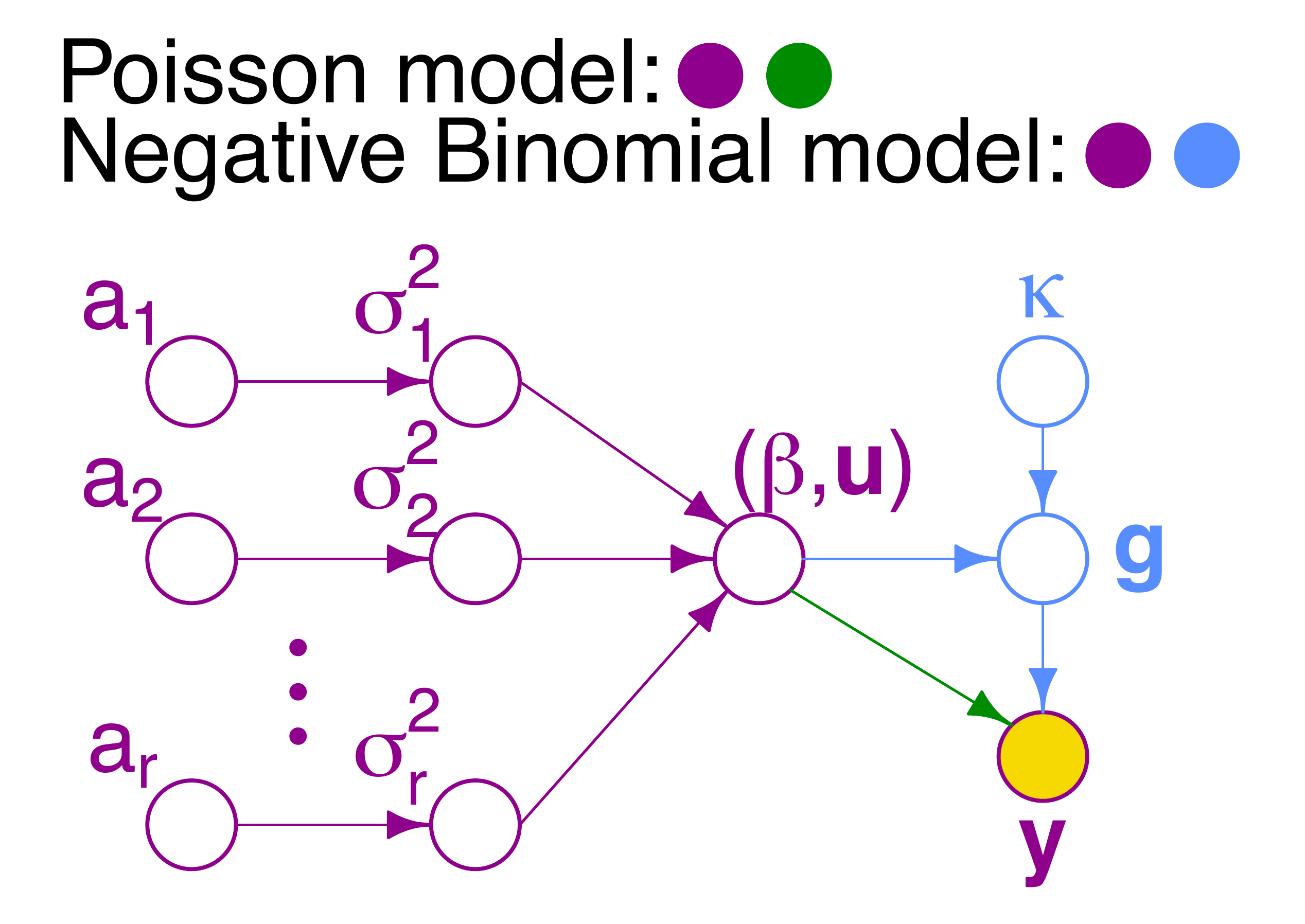}}
\caption{\it Directed acyclic graph corresponding to the models 
  (\ref{eq:poissonModel}) and (\ref{eq:negBinModel}). 
  The shaded node corresponds to the
  observed data. The color key at the top of the figure denotes
  the components of the graph corresponding to each model.}
\label{fig:dag} 
\end{figure}
}
{\vskip3mm
\thickboxit{\bf \centerline{dag figure here.}}
\vskip3mm
}

\subsection{Extension to Unstructured Covariance Matrices for Random Effects}

Section 2.3 of \citet{Luts13} describes the extension to semiparametric
models containing unstructured covariance matrices. Such extensions
arise in the case of random intercept and slope models. A simple
example of such a model having count responses is:
$$
\begin{array}{c}
y_{ij}|\beta_0,\beta_1,U_i,V_i\simind
\mbox{Poisson}\{\exp(\beta_0+U_i+(\beta_1+V_i)\,x_{ij})\},\quad 1\le i\le m,
\quad 1\le j\le n_i,\\[2ex]
\mbox{and}\quad
\left[
\begin{array}{c}
U_i\\
V_i \\
\end{array}
\right]\Big|\bSigma\sim N(\bzero,\bSigma),
\quad\mbox{where}\quad
\bSigma\equiv  \left[\begin{array}{cc}         
\sigma_u^2 & \rho_{uv}\,\sigma_u\,\sigma_v \\
\rho_{uv}\,\sigma_u\,\sigma_v & \sigma_v^2
\end{array}\right].
\end{array}
$$
The advice given in Section 2.3 of \citet{Luts13} 
concerning such extensions applies here as well.

\subsection{Hyperparameter Default Values}

With noninformativity in mind, reasonable default values for the hyperparameters
in models (\ref{eq:poissonModel}) and (\ref{eq:negBinModel})
are
$$\sigma_{\beta}=A_{\ell}=10^5,\quad 
\kappa_{\text{min}}=\textstyle{\frac{1}{100}}\quad\mbox{and}\quad
\kappa_{\text{max}}=100,$$
assuming that the predictor data have been standardized to have 
zero mean and unit standard deviation.

All examples in this article use these hyperparameter settings
with standardized predictor data, and then transform the
results to the original units.

\section{Variational Inference Scheme}\label{sec:VBinference}

We are now in a position to derive a variational inference scheme
for fitting the Poisson and Negative Binomial additive mixed models
described in Section \ref{sec:modelDescription} and 
displayed in Figure \ref{fig:dag}. In this section 
we work toward a variational inference algorithm  
that treats both models by taking advantage of their commonalities, 
but also recognizing the differences. The algorithm, which we call 
Algorithm \ref{alg:nbpBatchAlgorithmMFVB}, 
is given in Section \ref{sec:algoListing}.

\subsection{Poisson Case}\label{sec:VBpoisson}

We first treat the Poisson additive mixed model (\ref{eq:poissonModel}).
Ordinary MFVB begins with a product restriction such as 
\begin{equation}
p(\bbeta,\bu,\sigma_{1}^2,\ldots,\sigma_{r}^2,a_{1},\ldots,a_{r}|\by) 
\approx 
q(\bbeta,\bu)\,
\,q(\sigma^2_{1},\ldots,\sigma^2_{r})\,q(a_1,\ldots,a_r).
\label{eq:MFVBrestrict}
\end{equation}
However, under (\ref{eq:MFVBrestrict}), the optimal posterior density
function of $(\bbeta,\bu)$ is
$$q^*(\bbeta,\bu)\propto \exp[E_{q(-(\bbeta,\bu))}
\{\log p(\bbeta,\bu|\mbox{rest})\}]$$
and involves multivariate integrals that are
not available in closed form.
A non-conjugate variational message passing solution is 
one that instead works with
\begin{equation}
p(\bbeta,\bu,\sigma_{1}^2,\ldots,\sigma_{r}^2,a_{1},\ldots,a_{r}|\by) 
\approx 
q(\bbeta,\bu;\bmu_{q(\bbeta,\bu)},\bSigma_{q(\bbeta,\bu)})
\,q(\sigma^2_{1},\ldots,\sigma^2_{r})\,q(a_1,\ldots,a_r)
\label{eq:NCVMPrestrict}
\end{equation}
where 
\begin{equation}
q(\bbeta,\bu;\bmu_{q(\bbeta,\bu)},\bSigma_{q(\bbeta,\bu)})
\quad\mbox{is the}\quad N\left(\bmu_{q(\bbeta,\bu)},
\bSigma_{q(\bbeta,\bu)}\right)\quad \mbox{density function.}
\label{eq:MultivNormForq}
\end{equation}

In the appendix, we show that the optimal posterior densities for
the variance and auxiliary parameters are:
\begin{equation}
\begin{array}{l}
q^*(\sigma^2_1,\ldots,\sigma^2_r)\, \text{is the product of}\\[1ex] 
\mbox{Inverse-Gamma}\left( \frac{K_{\ell}+1}{2}, \mu_{q(1/a_{\ell})} + \frac{1}{2} 
\left\{\Vert \bmu_{q(\bu_{\ell})} \Vert^2 + \text{tr} 
\left( \bSigma_{q(\bu_{\ell})} \right) \right\} \right)
\ \mbox{density functions,}\\ [1ex]
\mbox{and}\ q^*(a_1,\ldots,a_r)\, \text{is the product of}\, \mbox{Inverse-Gamma} 
\left(1,\mu_{q(1/\sigma^{2}_{{\ell}})} + A^{-2}_{{\ell}} \right)\ \mbox{density}\\[1ex]
\text{functions, $1\le \ell\le r$,}
\end{array}
\label{eq:sigsqaOpt}
\end{equation}
where
$\mu_{q(1/\sigma_{\ell}^2)}\equiv 
\int_0^{\infty}(1/\sigma_{\ell}^2) q(\sigma_{\ell}^2)
\,d\sigma_{\ell}^2$,\ \ 
$\mu_{q(1/a_{\ell})}$ is defined analogously,
$$\bmu_{q(\bu_{\ell})}\equiv\mbox{sub-vector of $\bmu_{q(\bbeta,\bu)}$ corresponding
to $\bu_{\ell}$}$$
and 
$$\bSigma_{q(\bu_{\ell})}\equiv 
\mbox{sub-matrix of $\bSigma_{q(\bbeta,\bu)}$ corresponding
to $\bu_{\ell}$.}$$
The interdependencies between the parameters in these optimal density
functions, combined with the updates for $\bmu_{q(\bbeta,\bu)}$ 
and $\bSigma_{q(\bbeta,\bu)}$ given by (\ref{eq:updatesMatt})
give rise to an iterative scheme for their solution, and is
encompassed in Algorithm \ref{alg:nbpBatchAlgorithmMFVB}.

Algorithm \ref{alg:nbpBatchAlgorithmMFVB} also uses
the variational lower bound on the marginal log-likelihood.
For model (\ref{eq:poissonModel}) and restriction
(\ref{eq:NCVMPrestrict}) it has the explicit expression
\begin{equation}
\begin{array}{rcl}
\log\underline{p}(\by;q) & = & \frac{P}{2} - r \log (\pi) - \frac{p}{2} 
\log( \sigma^2_{\beta}) + \frac{1}{2} \log |\bSigma_{q(\bbeta,\bu)}| 
- \bone^T \log(\by!) \nonumber \\[2ex]
&& - \frac{1}{2 \sigma^2_{\beta}}\{\Vert\bmu_{q(\bbeta)}\Vert^2 
+ \text{tr}(\bSigma_{q(\bbeta)})\} +  
\sum_{\ell=1}^{r}\Big[\mu_{q(1/a_{\ell})} \mu_{q(1/\sigma_{\ell}^{2})} \\[2ex]
&& - \log(A_{\ell}) - \log\{\mu_{q(1/\sigma_{\ell}^{2})} 
+ A_{\ell}^{-2}\} + \log\left\{\Gamma \left( \frac{K_{\ell}+1}{2}\right)\right\} 
\\[2ex]
&&- \frac{K_{\ell}+1}{2} \log(\mu_{q(1/a_{\ell})} 
+ \frac{1}{2}\{\Vert \bmu_{q(\bu_{\ell})} \Vert^2 
+ \text{tr}( \bSigma_{q(\bu_{\ell})})\}\Big] \\[2ex]
&& + \by^T \bC \bmu_{q(\bbeta,\bu)} - \bone^T 
\exp \left\{ \bC \bmu_{q(\bbeta,\bu)} + \frac{1}{2} 
\mbox{diagonal}( \bC \bSigma_{q(\bbeta,\bu)} \bC^T)\right\}.\\
\end{array}
\end{equation}
Here and elsewhere, 
$$\mbox{diagonal}(\bM)\equiv\mbox{vector of diagonal entries of $\bM$}$$
for any square matrix $\bM$. Also,
$$\bC\equiv[\bX \ \bZ]\quad\mbox{and}\quad
P\equiv\mbox{number of columns in $\bC$}
=p+\sum_{\ell=1}^r K_{\ell}.
$$

\subsection{Negative Binomial Case}\label{sec:VBnegbin}

We now turn our attention to the Negative Binomial response 
semiparametric regression model (\ref{eq:negBinModel}) 
and posterior density function approximations of the form
$$
\begin{array}{l}
p(\bbeta,\bu,\bg,\kappa,\sigma_{1}^2,\ldots,\sigma_{r}^2,a_{1},\ldots,a_{r}|\by)\\[2ex]
\qquad
\approx q(\bbeta,\bu;\bmu_{q(\bbeta,\bu)},\bSigma_{q(\bbeta,\bu)})\,q(\bg)\,q(\kappa)\,
\,q(\sigma^2_{1},\ldots,\sigma^2_{r})\,q(a_1,\ldots,a_r)
\end{array}
$$
with $q(\bbeta,\bu;\bmu_{q(\bbeta,\bu)},\bSigma_{q(\bbeta,\bu)})$ 
given by (\ref{eq:MultivNormForq}).

The optimal $q$-density functions for $\sigma_1^2,\ldots,\sigma_r^2$
and $a_1,\ldots,a_r$ are given by (\ref{eq:sigsqaOpt}).
With $\bc_i$ denoting the $i$th row of $\bC$, the optimal densities
for $\bg$ and $\kappa$ are:

\begin{equation}
\begin{array}{l}
q^*(\bg)\, \text{is the product of}\\[1ex]
\mbox{Gamma} \left(\mu_{q(\kappa)}+y_i ,
1 + \mu_{q(\kappa)} \exp\left( -\bc_i^T \bmu_{q(\bbeta,\bu)} 
+ \frac{1}{2} \bc_i^T \bSigma_{q(\bbeta,\bu)} \bc_i \right)\right)\\[1ex]
\text{density functions over $1\le i\le n$ and}\\[1ex]
q^*(\kappa)\, =\displaystyle{\frac{\exp[ n \left\{\kappa \log \left( \kappa \right) 
- \log \left( \Gamma\left(\kappa\right) \right) \right\} 
- C_1 \kappa ]}{\Hsc(0,n,C_1,\kappa_{\text{min}},\kappa_{\text{max}})}}, 
\quad \kappa_{\text{min}} \leq \kappa \leq \kappa_{\text{max}},
\end{array}
\label{eq:nbOptimalQ}
\end{equation}
where $\mu_{q(\kappa)}\equiv\int_{\kappa_{\text{min}}}^{\kappa_{\text{max}}}
\,\kappa\,q(\kappa)\,d\kappa$,
\begin{equation}
\Hsc(p,q,r,s,t)\equiv\int_s^t x^p 
\exp\Big(q [x \log(x) - \log\{\Gamma(x)\}]- r\, x\Big)\,dx, 
\quad p \geq 0, \quad q,r,s,t >0,
\label{eq:HscDefn}
\end{equation}
and 
$$C_1\equiv 
\bone^T \bC \bmu_{q(\bbeta,\bu)} - \bone^T \bmu_{q(\log(\bg))} 
+ \bmu_{q(\bg)}^T \exp \{ -\bC \bmu_{q(\bbeta,\bu)} + \frac{1}{2} \, 
\mbox{diagonal}( \bC \bSigma_{q(\bbeta,\bu)} \bC^T) \}.
$$
Details on the derivation of (\ref{eq:nbOptimalQ}) are given
in the appendix.

Algorithm \ref{alg:nbpBatchAlgorithmMFVB} provides an iterative scheme
for obtaining all $q$-density parameters. The marginal log-likelihood
lower-bound for the Negative Binomial case is
\begin{equation}
\begin{array}{rcl}
\log\underline{p}(\by;q) & = & \frac{P}{2} - r \log (\pi) -
\frac{p}{2}  \log( \sigma^2_{\beta}) + \frac{1}{2} \log
|\bSigma_{q(\bbeta,\bu)}| 
- \frac{1}{2 \sigma^2_{\beta}} \left( \Vert\bmu_{q(\bbeta)}\Vert^2 + 
\text{tr} \left( \bSigma_{q(\bbeta)} \right) \right) \nonumber \\[2ex]
&& + \sum_{l=1}^{r}\Big( \mu_{q(1/a_{\ell})}
  \mu_{q(1/\sigma_{\ell}^{2})} 
- \log \left( A_{\ell} \right) - \log \left\{
  \mu_{q(1/\sigma_{\ell}^{2})} 
+ A_{\ell}^{-2} \right\}  \\[2ex]
&& - \frac{K_{\ell}+1}{2} \log \left[ \mu_{q(1/a_{\ell})} 
+ \frac{1}{2} \left\{\Vert \bmu_{q(\bu_{\ell})} \Vert^2 + 
\text{tr}(\bSigma_{q(\bu_{\ell})}) \right\} \right] 
+ \log \left\{\Gamma\left(\frac{K_{\ell}+1}{2}\right)\right\}\Big)  \\[2ex]
&&  + \bone^T \log\left\{\Gamma\left( \mu_{q(\kappa)}\bone + \by
  \right)\right\} 
- \mu_{q(\kappa)} \bone^T \bmu_{q(\log(\bg))} - \bone^T\log(\by!)  \\[2ex]
&&  - (\by+\mu_{q(\kappa)}\bone)^T \log\left[\bone + \mu_{q(\kappa)} 
\exp\left\{ -\bC \bmu_{q(\bbeta,\bu)} + \frac{1}{2} \, 
\mbox{diagonal}(\bC \bSigma_{q(\bbeta,\bu)} \bC^T) \right\} \right] \\[2ex]
&& + \mu_{q(\kappa)} \bmu^T_{q(\bg)} 
\exp\left\{ -\bC\bmu_{q(\bbeta,\bu)} 
+ \frac{1}{2} \, \mbox{diagonal}( \bC \bSigma_{q(\bbeta,\bu)} \bC^T) 
\right\}\\[2ex]
&&- \log\left( \kappa_{\text{max}} - \kappa_{\text{min}} \right) 
+ \log\left\{\Hsc(0,n,C_1,\kappa_{\text{min}},\kappa_{\text{max}}) \right\}.\\
\null\\
\end{array}
\end{equation}

\subsection{Algorithm}\label{sec:algoListing}

We now present Algorithm \ref{alg:nbpBatchAlgorithmMFVB}. 
Note that $\bA\odot\bB$ denotes the element-wise 
product of two equal-sized matrices $\bA$ and $\bB$. 
Function evaluation is also interpreted in
an element-wise fashion. For example 
$$\Gamma\left(\left[\begin{array}{c}
7\\
3\\
9
\end{array}\right]\right)
\equiv
\left[\begin{array}{c}
\Gamma(7)\\
\Gamma(3)\\
\Gamma(9)
\end{array}\right].
$$
The digamma function is given by $\mbox{digamma}(x)\equiv 
\frac{d}{dx}\log\{\Gamma(x)\}$.
Most of the updates in Algorithm \ref{alg:nbpBatchAlgorithmMFVB}
require standard arithmetic. The exception is the function
$\Hsc$ defined by (\ref{eq:HscDefn}), and it is evaluated
using efficient quadrature strategies as described in 
Appendix B of \citet{Wand11}.

\begin{algorithm}
\begin{center}
\begin{minipage}[t]{161mm}
\hrule
\begin{itemize}
\item[] Initialize:\ $\mu_{q(1/\sigma_{\ell}^2)}>0 \, 
(1\leq \ell \leq r), \mu_{q(\kappa)}, \bmu_{q(\bbeta,\bu)}$ 
a $P \times 1$ vector and $\bSigma_{q(\bbeta,\bu)}$ a $P \times P$ 
positive definite matrix.
\item[] Cycle:
\begin{itemize}
\item[] $\bM_{q(1/\bsigma^2)} \leftarrow 
\mbox{blockdiag}(\sigma_{\beta}^{-2}\bI_p,\mu_{q(1/\sigma_1^2)}
\bI_{K_1},\ldots,\mu_{q(1/\sigma_r^2)}\bI_{K_r})$

\item[] If fitting the Poisson response model (\ref{eq:poissonModel}):
\begin{itemize}
\item[] \quad $\bw_{q(\bbeta,\bu)} \leftarrow \exp\{\bC\bmu_{q(\bbeta,\bu)}
+\smhalf\mbox{diagonal}(\bC\bSigma_{q(\bbeta,\bu)}\bC^T)\}$
\item[] \quad $\bmu_{q(\bbeta,\bu)} \leftarrow \bmu_{q(\bbeta,\bu)} 
+ \bSigma_{q(\bbeta,\bu)} 
\left\{\bC^T\left(\,\by-\bw_{q(\bbeta,\bu)}\right)-\bM_{q(1/\bsigma^2)}
\bmu_{q(\bbeta,\bu)} \right\}$
\item[] \quad $\bmu_{q(\bg)} \leftarrow \bone$; \quad $\mu_{q(\kappa)} 
\leftarrow 1$
\end{itemize}

\item[] If fitting the Negative Binomial response model (\ref{eq:negBinModel}):
\begin{itemize}
\item[] \quad $\bw_{q(\bbeta,\bu)} \leftarrow 
\exp\{-\bC\bmu_{q(\bbeta,\bu)}
+\smhalf\mbox{diagonal}(\bC\bSigma_{q(\bbeta,\bu)}\bC^T)\}$
\item[] \quad $\bmu_{q(\bg)} \leftarrow
(\mu_{q(\kappa)}\bone+\by)/(\bone+\mu_{q(\kappa)}\bw_{q(\bbeta,\bu)})$
\item[] \quad $\bmu_{q(\bbeta,\bu)} \leftarrow \bmu_{q(\bbeta,\bu)} 
+ \bSigma_{q(\bbeta,\bu)} \left\{ \mu_{q(\kappa)} \bC^T\left(\,\bmu_{q(\bg)} 
\odot \bw_{q(\bbeta,\bu)}-\bone \right)-\bM_{q(1/\bsigma^2)}
\bmu_{q(\bbeta,\bu)} \right\} $
\item[] \quad $\bmu_{q(\log(\bg))} \leftarrow 
\mbox{digamma}(\bone\mu_{q(\kappa)}+\by)
-\log(\bone+\mu_{q(\kappa)}\bw_{q(\bbeta,\bu)})$
\item[] \quad $C_1 \leftarrow  \bone^T \bC \bmu_{q(\bbeta,\bu)} 
- \bone^T \bmu_{q(\log(\bg))} + \bmu^T_{q(\bg)} \bw_{q(\bbeta,\bu)}$
\item[] \quad $\mu_{q(\kappa)} \leftarrow 
\exp\left[\log\left\{\Hsc(1,n,C_1,\kappa_{\text{min}},
\kappa_{\text{max}})\right\} - 
\log\left\{\Hsc(0,n,C_1,
\kappa_{\text{min}},\kappa_{\text{max}})\right\}\right]$
\end{itemize}
\item[] $\bSigma_{q(\bbeta,\bu)} \leftarrow 
\left\{ \mu_{q(\kappa)} \bC^T\diag(\bmu_{q(\bg)}\odot 
\bw_{q(\bbeta,\bu)})\,\bC+\bM_{q(1/\bsigma^2)}\right\}^{-1}$
\item[] For $\ell = 1,\ldots,r:$
\item[] \quad $\mu_{q(1/a_{\ell})}\leftarrow 
1/\{\mu_{q(1/\sigma_{\ell}^2)}+A_{\ell}^{-2}\}$\ \  ;\ \  
$\mu_{q(1/\sigma_{\ell}^2)}\leftarrow 
\displaystyle{\frac{K_{\ell}+1}{2\,\mu_{q(1/a_{\ell})}
+\Vert\bmu_{q(\bu_{\ell})}\Vert^2+\mbox{tr}(\bSigma_{q(\bu_{\ell})})}}$%
\end{itemize}
\item[] until the relative change in $\punder(\by;q)$ is negligible.
\end{itemize}
\hrule
\end{minipage}
\end{center}
\caption{\it Non-conjugate MFVB algorithm for approximate inference in either the 
Poisson response model (\ref{eq:poissonModel}) or 
the Negative Binomial response model (\ref{eq:negBinModel}).}
\label{alg:nbpBatchAlgorithmMFVB} 
\end{algorithm}

\null\vfill\eject

\section{Real-time Count Response Semiparametric Regression}\label{sec:real-time}

An advantage of MFVB approaches to approximate inference is their adaptability
to real-time processing. As discussed in Section \ref{sec:intro}, this is important 
for both high volume and/or velocity data. Here we briefly present
an adaptation of the Poisson component of Algorithm \ref{alg:nbpBatchAlgorithmMFVB} 
that permits real-time count response semiparametric regression.

Rather than processing $\by$ and $\bC$ in batch,
as done by Algorithm \ref{alg:nbpBatchAlgorithmMFVB}, Algorithm
\ref{alg:nbpOnlineAlgorithmMFVB} processes each new entry 
of $\by$, denoted by $\ynew$, and its corresponding row of $\bC$, 
denoted by $\bcnew$, sequentially in real time. 

\citet{Luts13} stress the importance of batch runs for 
determination of starting values for real-time semiparametric
regression procedures and their Algorithm 2' formalized 
such a strategy. This is reflected in Algorithm
\ref{alg:nbpOnlineAlgorithmMFVB}. We also found it
necessary to not use the value of $\mu_{q(\bbeta,\bu)}$
from the previous iteration in its update but, rather,
a value from a previous iteration. The turning parameter
$\updateMult>1$ controls the rate at which 
previous versions of $\mu_{q(\bbeta,\bu)}$ are used
in its update, and a reasonable default is $\updateMult=100$.

\begin{algorithm}
\begin{center}
\begin{minipage}[t]{150mm}
\hrule
\begin{itemize}
\item[1.] Use Algorithm \ref{alg:nbpBatchAlgorithmMFVB} 
to perform batch-based tuning runs, 
analogous to those described in Algorithm 2' of \citet{Luts13},
and determine a warm-up sample size $\nwarm$ for which
convergence is validated. 
\item[2.] Set 
$\bmu_{q(\bbeta,\bu)},\mu_{q(1/\sigma_1^2)},\ldots,
\mu_{q(1/\sigma_r^2)}$ and $w_{q(\bbeta,\bu)}$
to be the values for these quantities obtained in the 
batch-based tuning run with sample size $\nwarm$.
Then set $\bywarm$ and $\bCwarm$ to be the response vector
and design matrix based on the first $\nwarm$ observations
and put
\ $\bC^T\,\by\leftarrow \bCwarm^T\bywarm$,
\ $\bC^T\,\bw_{q(\bbeta,\bu)}\leftarrow \bCwarm^T\bw_{q(\bbeta,\bu)}$,
\ $\bC^T\diag(\bw_{q(\bbeta,\bu)})\bC\leftarrow
\bCwarm^T\diag(\bw_{q(\bbeta,\bu)})\bCwarm$,
\ $n\leftarrow\nwarm$. 
Lastly, set $\bmu_{\text{prev}} \leftarrow \bmu_{q(\bbeta,\bu)}$
and $\updateMult>1$ to be an integer (defaulted
to be $\updateMult=100$).
\item[3.] Cycle:
\begin{itemize}
\item[] Read in $y_{\text{new}}\ (1\times 1)$ and 
$\bc_{\text{new}}\ (P\times 1) \quad ; \quad n 
\leftarrow n + 1$
\item[] $\bM_{q(1/\bsigma^2)} \leftarrow 
\mbox{blockdiag}(\sigma_{\beta}^{-2}\bI_p,\mu_{q(1/\sigma_1^2)}\bI_{K_1},
\ldots,\mu_{q(1/\sigma_r^2)}\bI_{K_r})$
\item[] $\bw_{q(\bbeta,\bu)} \leftarrow \exp(\bc_{\text{new}}^T 
\bmu_{q(\bbeta,\bu)}+\smhalf \bc_{\text{new}}^T\bSigma_{q(\bbeta,\bu)}
\bc_{\text{new}})$
\item[] $\bC^T\by \leftarrow 
\bC^T\by  + \bc_{\text{new}} y_{\text{new}}$
\ \ \ \ ;\ \ \  
$\bC^T \bw_{q(\bbeta,\bu)} \leftarrow 
\bC^T \bw_{q(\bbeta,\bu)} + \bc_{\text{new}}w_{q(\bbeta,\bu)}$
\item[] $\bC^T\diag(\bw_{q(\bbeta,\bu)})\,\bC \leftarrow 
\bC^T\diag(\bw_{q(\bbeta,\bu)})\,\bC + w_{q(\bbeta,\bu)} 
\bc_{\text{new}} \bc_{\text{new}}^T$
\item[] $\bmu_{q(\bbeta,\bu)} \leftarrow \bmu_{\text{prev}} 
+ \bSigma_{q(\bbeta,\bu)} \left\{ \bC^T\by - \bC^T\bw_{q(\bbeta,\bu)}
-\bM_{q(1/\bsigma^2)}
\bmu_{q(\bbeta,\bu)}\right\}$
\item[] If $n$ is a multiple of $\updateMult$ then 
$\bmu_{\text{prev}} \leftarrow \bmu_{q(\bbeta,\bu)}$.
\item[] $\bSigma_{q(\bbeta,\bu)} \leftarrow 
\left\{\bC^T\diag(\bw_{q(\bbeta,\bu)})\,\bC+\bM_{q(1/\bsigma^2)}\right\}^{-1}$
\item[] For $\ell = 1,\ldots,r:$
\item[] \quad $\mu_{q(1/a_{\ell})}\leftarrow
  1/\{\mu_{q(1/\sigma_{\ell}^2)}
+A_{\ell}^{-2}\}$
\item[]
\quad $\mu_{q(1/\sigma_{\ell}^2)}\leftarrow 
\displaystyle{\frac{K_{\ell}+1}{2\,\mu_{q(1/a_{\ell})}
+\Vert\bmu_{q(\bu_{\ell})}\Vert^2+\mbox{tr}(\bSigma_{q(\bu_{\ell})})}}$%
\end{itemize}
\item[] until data no longer available or analysis terminated.
\end{itemize}
\hrule
\end{minipage}
\end{center}
\caption{\it Online non-conjugate variational message
passing algorithm for real-time approximate inference 
in the Poisson response model (\ref{eq:poissonModel}).}
\label{alg:nbpOnlineAlgorithmMFVB} 
\end{algorithm}

An illustration of Algorithm \ref{alg:nbpOnlineAlgorithmMFVB}
is described in Section \ref{sec:movie}.

\section{Numerical Results}\label{sec:numerical}

Algorithms \ref{alg:nbpBatchAlgorithmMFVB} and \ref{alg:nbpOnlineAlgorithmMFVB}
have been tested on various synthetic and actual data-sets. 
We first describe the results
of a simulation study that allows us to make some summaries
of the accuracy of MFVB in this context. This is followed
by some applications. Lastly, we describe an illustration
of Algorithm \ref{alg:nbpOnlineAlgorithmMFVB} that takes
the form of a movie on our real-time semiparametric regression 
web-site.

\subsection{Simulation Study}

We ran a simulation study involving the true mean function
\begin{equation}
\begin{array}{l}
\mu(x_{1},x_{2})\equiv\exp\{g_1(x_{1})+g_2(x_{2})\}\nonumber
\end{array}
\label{eq:modelForsimulation}
\end{equation}
where
\begin{eqnarray*}
g_1(x)&\equiv&\cos(4\pi x)\,+2\,x,\\[2ex] 
g_2(x)&\equiv&0.4 \,\phi(x;0.38,0.08) -
1.02 \, x + 0.018 \, x^2 + 0.08 \, \phi(x;0.75,0.03)
\end{eqnarray*} 
and $\phi(x;\mu,\sigma)$ 
denotes the value of the Normal density function with mean
$\mu$ and standard deviation $\sigma$ evaluated at $x$. 
Next, we generated 100 data-sets, each having
500 triplets $(y_i,x_{1i},x_{2i})$, using the Poisson response model
\begin{equation}
y_i \simind \mbox{Poisson}(\mu(x_{1i},x_{2i})),\quad 1 \leq i \leq 500,
\label{poissonSim}
\end{equation}
and the Negative Binomial response model
\begin{equation}
	y_i \simind \mbox{Negative-Binomial}(\mu(x_{1i},x_{2i}),3.8),\quad 1 \leq i \leq 500,
	\label{nbSim}
\end{equation}
where $x_{1i},x_{2i}\simind \mbox{Uniform}(0,1)$.
We model  $g_1(x_{1})+g_2(x_{2})$ using mixed model-based
penalized splines \citep[e.g.][]{Ruppert03}: 
\begin{equation}
\begin{array}{c}
	\beta_0+\beta_1\,x_{1}+\beta_2\,x_{2}+\sum_{k=1}^{K_1} 
         u_{1k}\,z_{1k}(x_{1})+\sum_{k=1}^{K_2} u_{2k}\,z_{2k}(x_{2}),\\[2ex]
u_{1k}|\,\sigma_1^2\simind N(0,\sigma_1^2),\quad 
u_{2k}|\,\sigma_2^2\simind N(0,\sigma_2^2),
\end{array}
\label{eq:OSull}	
\end{equation}
where $z_{1k}$ and $z_{2k}$ represent O'Sullivan 
splines \citep{Wand08}. After grouping 
$\bbeta = [\beta_0 \, \beta_1 \, \beta_2]^T$, 
$\bu = [u_{11}, \ldots, u_{1{K_1}}, u_{21}, \ldots, u_{2{K_2}}]^T$ 
and creating the corresponding design matrices $\bX$ and $\bZ$,
Algorithm \ref{alg:nbpBatchAlgorithmMFVB} is used for MFVB inference. 
We set the number of spline basis functions to be $K_1 = K_2 = 17$. 
The MFVB iterations were terminated
when the relative change in $\log\underline{p}(\by;q)$ was less than 
$10^{-10}$.

For MCMC analysis 5000 samples were generated after a burn-in of size
5000. Thinning with a factor of 5 resulted in 1000 retained MCMC
samples for inference. MCMC analysis was performed in \proglang{BUGS}. 

\subsubsection{Accuracy assessment}

Figure \ref{fig:pAccuracies} displays side-by-side boxplots of the
accuracy scores for the parameters in the Poisson response
simulation study. For a generic parameter $\theta$, the
accuracy score is defined by
\begin{equation}
	\text{accuracy}(q^*) = 100 \left( 1- \frac{1}{2} 
\int^{\infty}_{-\infty} |q^*(\theta) -p(\theta|\by)| \, d\theta
\right)\%.
\nonumber
\end{equation}
Note that a kernel
density estimate based on the MCMC samples is used for the posterior 
density function $p(\theta|\by)$.

\ifthenelse{\boolean{ShowFigures}}
{
\begin{figure}[!ht]
\centering
{\includegraphics[width=\textwidth]{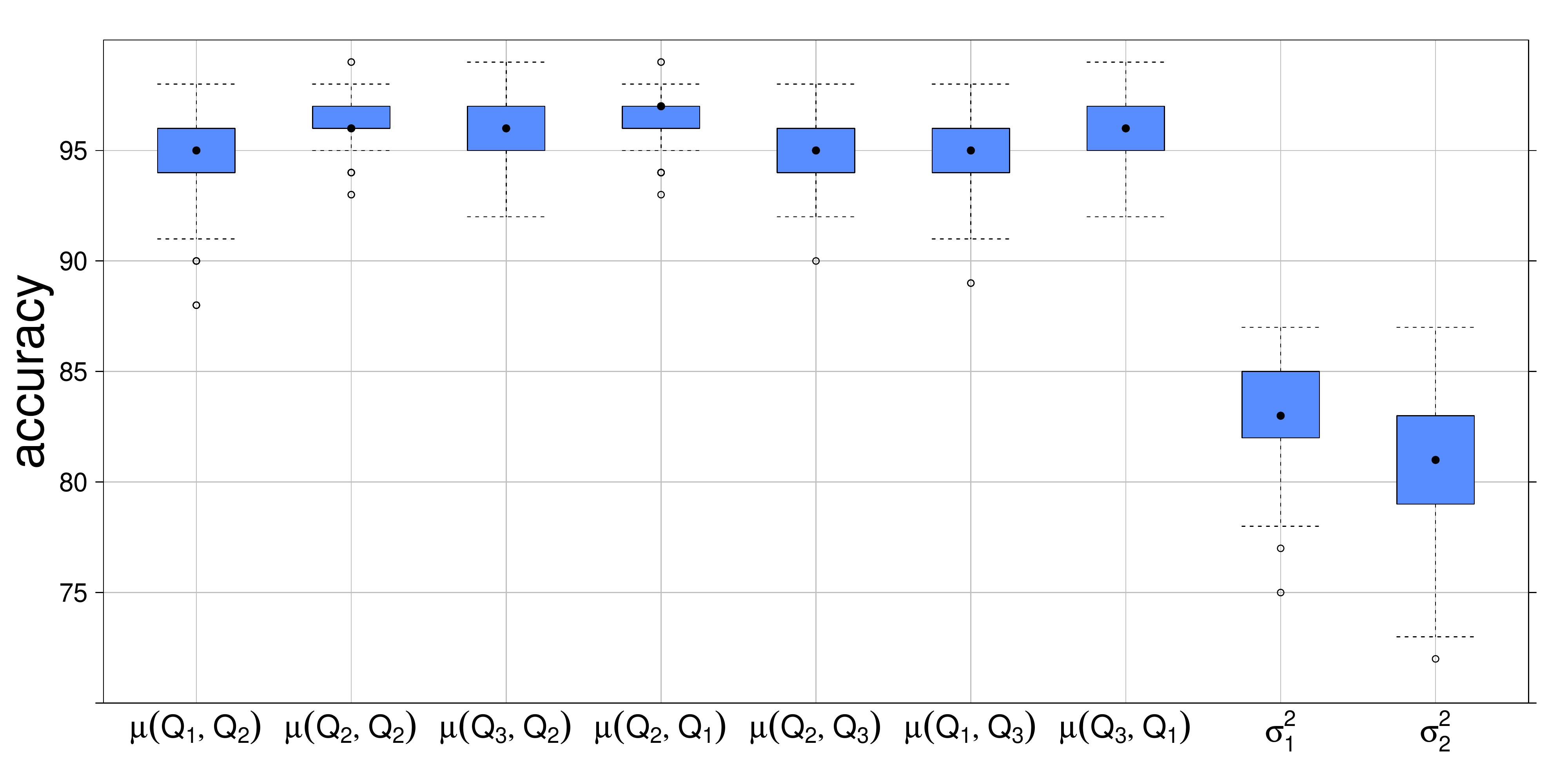}}
\caption{\it Side-by-side boxplots of accuracy values for MFVB against
 an MCMC benchmark for the Poisson response model (\ref{poissonSim}).}
\label{fig:pAccuracies} 
\end{figure}
}
{\vskip3mm
\thickboxit{\bf \centerline{Poisson MFVB versus MCMC accuracies.}}
\vskip3mm
}

The parameters on the horizontal axis of Figure \ref{fig:pAccuracies}
represent the estimated approximate posterior density functions for 
$\mu(x_{1},x_{2})$, evaluated at the quartiles of $x_{1}$ and $x_{2}$,
and the estimated approximate posterior density functions for
$\sigma_1^2$ and $\sigma_2^2$. The boxplots indicate that the
accuracies for $\mu(x_{1},x_{2})$ are around 95\%, while values
between 80\% and  85\% are obtained for the variances $\sigma_1^2$ and
$\sigma_2^2$.

Figure \ref{fig:poissonApproxDist} shows the MFVB-based approximate
posterior density functions against the MCMC result for a single
replicated data-set. The accuracy of MFVB is particularly excellent
for the $\mu(x_{1},x_{2})$ approximate posterior density functions.

\ifthenelse{\boolean{ShowFigures}}
{
\begin{figure}[!ht]
\centering
{\includegraphics[width=\textwidth]{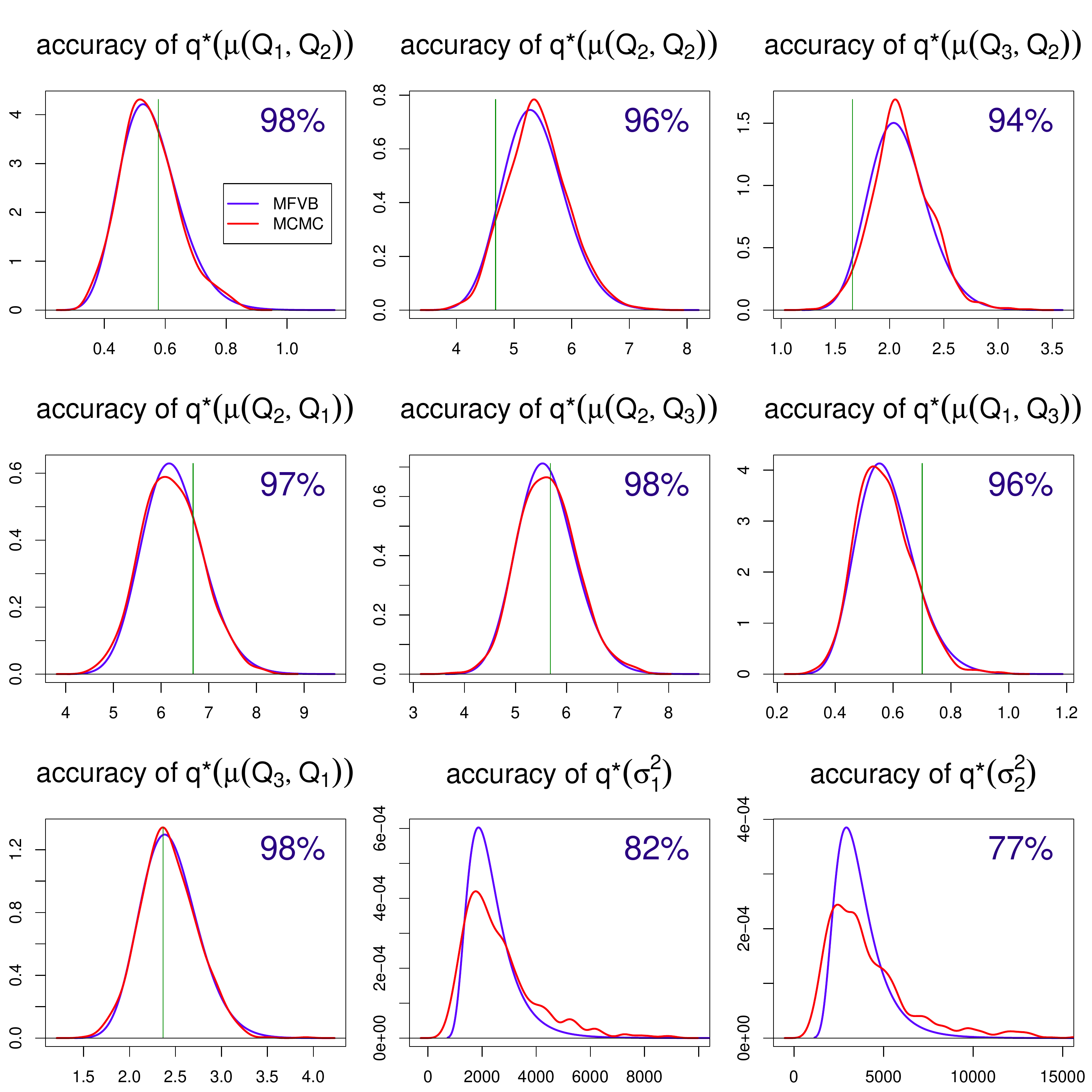}}
\caption{\it Approximate posterior density functions for Poisson
  response model (\ref{poissonSim}). Vertical lines indicate the true
  values.}
\label{fig:poissonApproxDist} 
\end{figure}
} {\vskip3mm \thickboxit{\bf \centerline{Poisson approximate posterior
      density functions figure here.}}  \vskip3mm }

Figure \ref{fig:negBinAccuracies} displays side-by-side boxplots of
the accuracies for the 100 data-sets generated according to the
Negative Binomial response model (\ref{nbSim}).

\ifthenelse{\boolean{ShowFigures}}
{
\begin{figure}[!ht]
\centering
{\includegraphics[width=\textwidth]{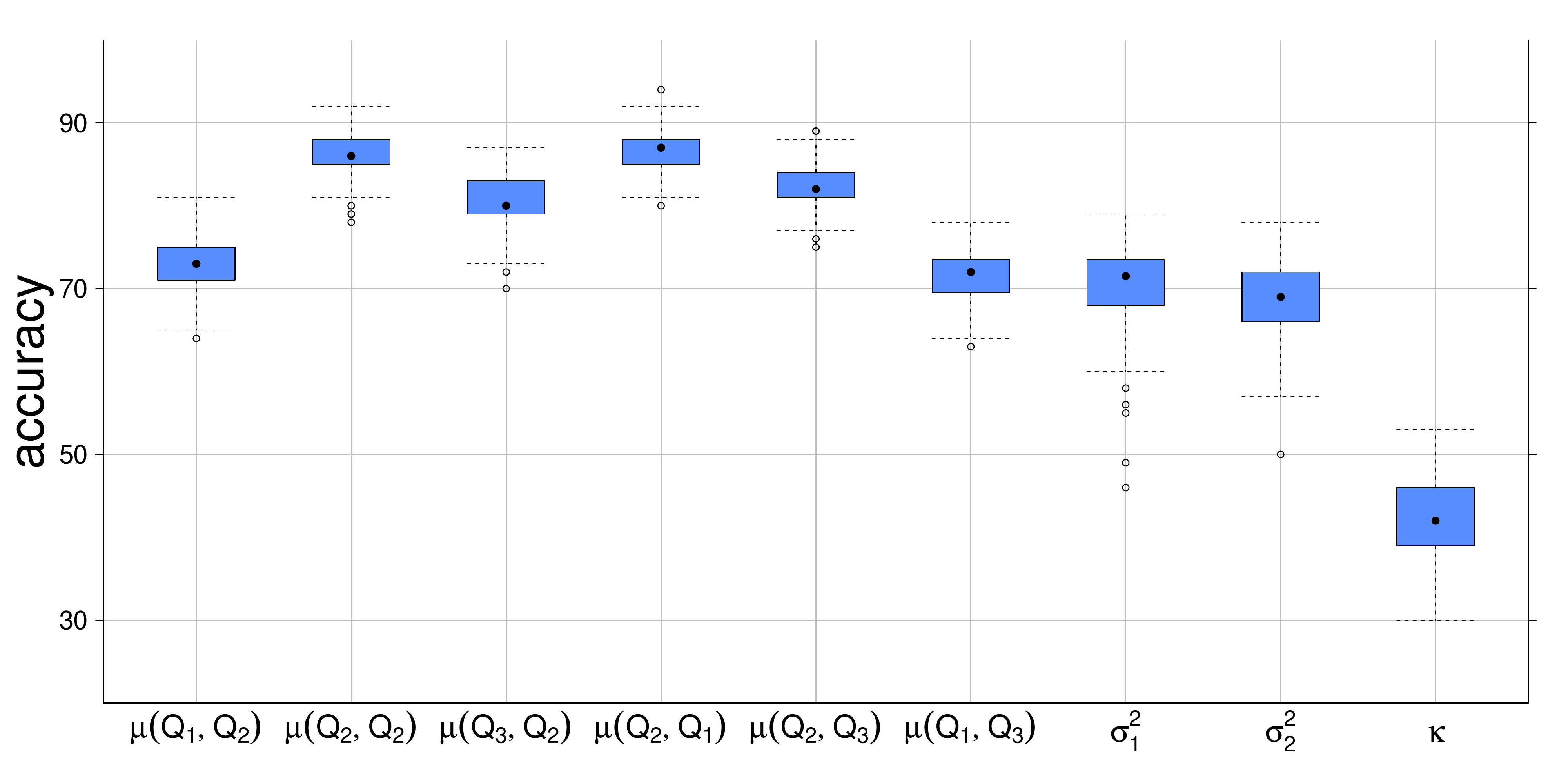}}
\caption{\it Side-by-side boxplots of accuracy values for MFVB against an 
MCMC benchmark for Negative Binomial response model (\ref{nbSim}).}
\label{fig:negBinAccuracies} 
\end{figure}
}
{\vskip3mm
\thickboxit{\bf \centerline{Negative Binomial MFVB versus MCMC accuracies.}}
\vskip3mm
}

The parameters on the horizontal axis in Figure \ref{fig:negBinAccuracies} 
have similar meanings as in Figure \ref{fig:pAccuracies}, but the result 
for the approximate posterior density function of $\kappa$ is also included. 
Compared to the results for the Poisson case the accuracies for the Negative 
Binomial response model are lower, but still attain good performance for 
$\mu(x_{1},x_{2})$ with approximately values between 70 and 90\%. The 
majority of the accuracies for the variances $\sigma_1^2$ and $\sigma_2^2$ 
is around 70\%, while lower accuracies are obtained for $\kappa$.

Finally, Figure \ref{fig:negBinApproxDist} compares the approximate
posterior density functions obtained using MFVB inference and the MCMC
result for a single replicated data-set. MFVB attains particularly
good accuracies for the $\mu(x_{1},x_{2})$ approximate posterior
density functions.

\ifthenelse{\boolean{ShowFigures}}
{
\begin{figure}[!ht]
\centering
{\includegraphics[width=\textwidth]{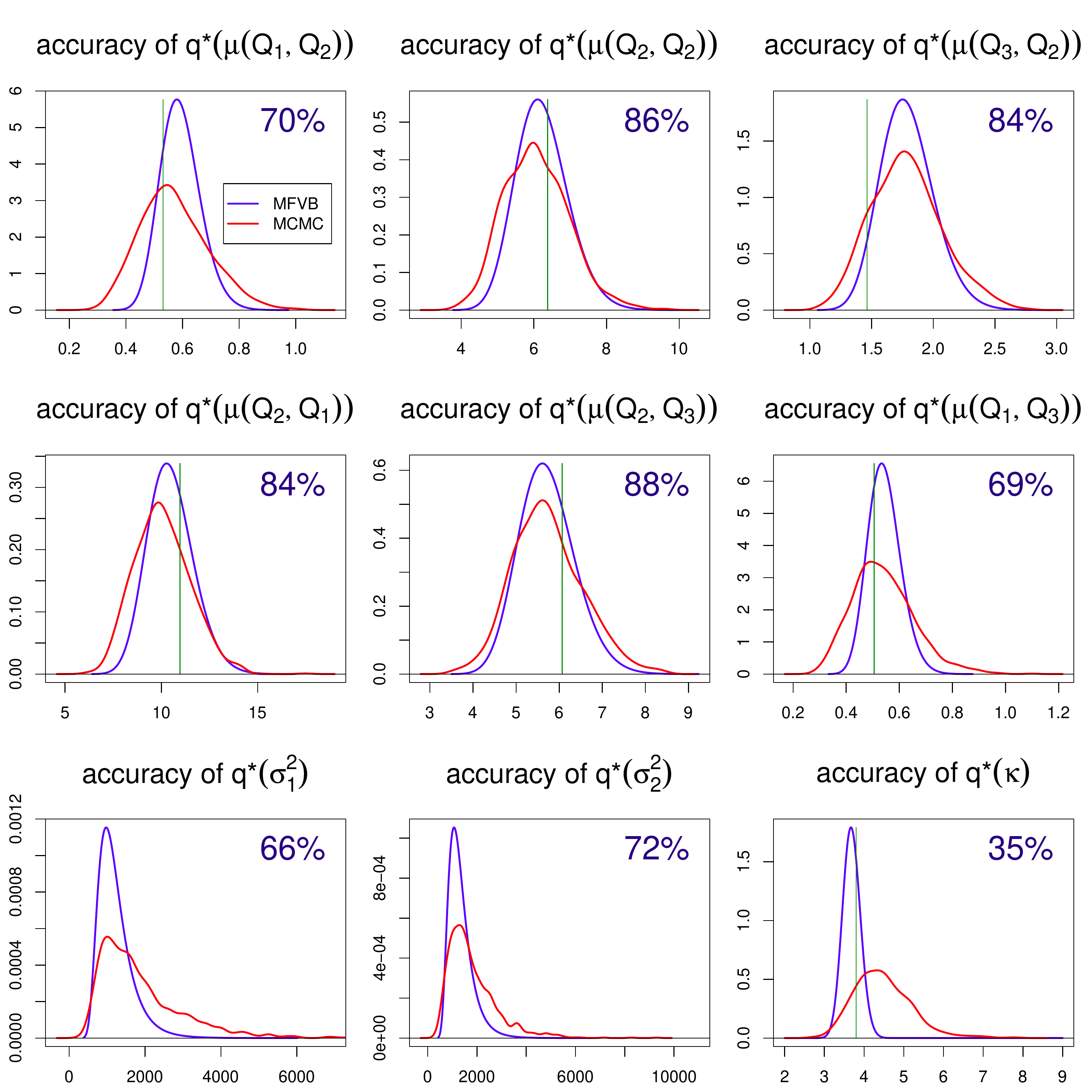}}
\caption{\it Approximate posterior density functions for Negative
  Binomial response model (\ref{nbSim}). Vertical lines indicate the
  true values.}
\label{fig:negBinApproxDist} 
\end{figure}
} {\vskip3mm \thickboxit{\bf \centerline{Negative Binomial approximate
      posterior density functions figure here.}}  \vskip3mm }

\subsubsection{Computational cost}

Table \ref{computationCost} summarizes the computation times for MCMC
and MFVB 
fitting in case of the Poisson and Negative Binomial
experiment as run using an Intel Core i7-2760QM  2.40 GHz processor
with 8 GBytes of random access memory. The average computing time for MFVB is
considerably lower than that of MCMC. Nevertheless, the speed
gains of MFVB need to be traded off against accuracy losses as shown
in Figures \ref{fig:pAccuracies} and \ref{fig:negBinAccuracies}.

\begin{table}[h!!]
\begin{center}
\begin{tabular}{lcc}
& MCMC & MFVB \\[0.1ex]
\hline\\[-0.9ex]
Poisson response model &856.66 (23.13) &2.24 (0.30)  \\
Negative Binomial response model  &1127.96 (56.73) & 20.85 (3.52) \\[1ex]
\hline
\end{tabular}
\caption{\it Average (standard deviation) times in seconds for MCMC
 and MFVB inference based on the simulation study.}
\label{computationCost}
\end{center}
\end{table}

\subsection{Applications}

We now present some applications involving
each of models (\ref{eq:poissonModel}) and (\ref{eq:negBinModel})
in turn.

\subsubsection{North African Conflict}

We fitted the Poisson response model (\ref{eq:poissonModel}) 
using Algorithm \ref{alg:nbpBatchAlgorithmMFVB} to a data-set 
extracted from the Global Database of Events,
Language and Tone \citep{Leetaru13}. This database contains
more than 200 million geo-located events, obtained from news reports,
with global coverage between early 1979 and June 2012. For this example we
extracted the daily number of material conflicts for each
African country for the period September 2010 to June 2012. Our model is
\begin{equation}
\texttt{conflicts}_{ij}|\,\bbeta,\bu_1,U_i \simind \mbox{Poisson}
(\exp\{ \beta_0 + f_1(\texttt{time}_{j}) + U_i \}), \nonumber
\end{equation}
with $\texttt{conflicts}_{ij}$ the number of news reports about
material conflicts for country $i$ on date $j$, $\texttt{time}_j$ the
time in days for date $j$ starting from September 1, 2010 and $U_i$
the random intercept for country $i$, $1 \leq i \leq 54$. The total
number of observations for all African countries is $n=36126$.
Note that 20 spline basis functions were used for modeling $f_1$.

Figure \ref{fig:poissonAfrica} shows the estimate for $\exp\{ \beta_0
+ f_1(\texttt{time}_{j})\}$ and corresponding 95\% pointwise credible
sets. The strong increase, starting around December 2010, in number of
news reports about material conflicts coincides with the Arab Spring
demonstrations and civil wars which took place in several African
countries as Mauritania, Western Sahara, Morocco, Algeria, Tunisia,
Libya, Egypt, Sudan, Djibouti and the related crisis in Mali. In
addition, 95\% credible sets for the estimates of $\exp(U_i)$ are
plotted for the fifteen countries with the largest random intercept
estimates, i.e. showing larger numbers of material conflict-related
news reports. Fitting using Algorithm \ref{alg:nbpBatchAlgorithmMFVB}
took $7$ minutes and $30$ seconds.

\ifthenelse{\boolean{ShowFigures}}
{
\begin{figure}[!ht]
\centering
{\includegraphics[width=\textwidth]{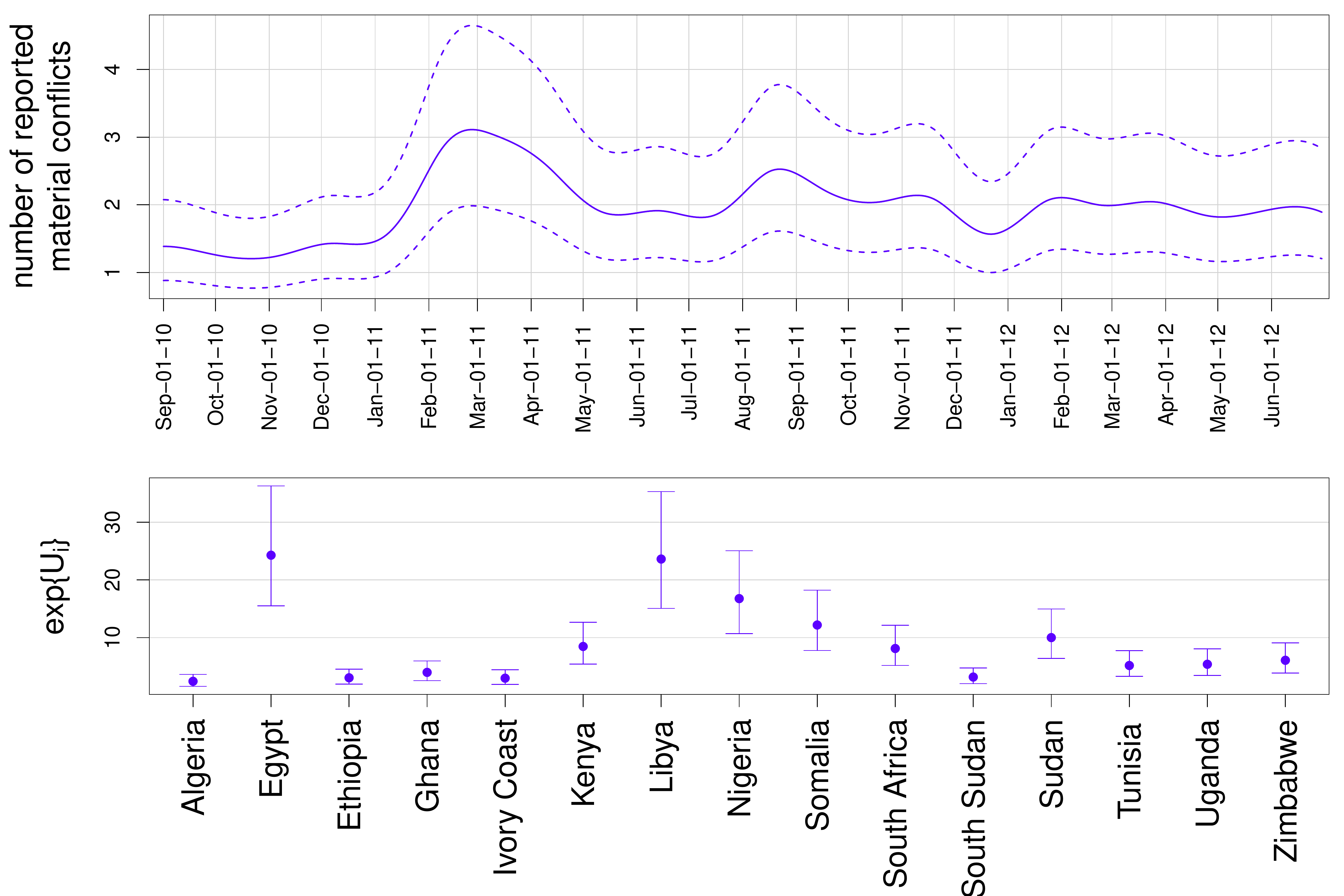}}
\caption{\it Poisson regression result using MFVB inference for global
data on events, location and tone database. The solid curve in the
top panel are posterior means and the dashed curves are pointwise
95\% credible sets. The lower panel shows 95\% credible sets for the 
estimates of $\exp(U_i)$ for the fifteen countries with highest posterior means.}
\label{fig:poissonAfrica} 
\end{figure}
}
{\vskip3mm
\thickboxit{\bf \centerline{Negative Binomial result for GDELT data figure here.}}
\vskip3mm
}

\subsubsection{Adduct data}

Illustrations of Negative Binomial semiparametric regression
models have previously been given in \citet{Thurston00}
and \citet{Marley10} using data on adducts counts, which are
carcinogen-DNA complexes, and smoking variables for 
78 former smokers in the lung cancer study \citep{Wiencke99}.
Here we use Algorithm \ref{alg:nbpBatchAlgorithmMFVB} to fit a 
version of the Bayesian penalized model that \citet{Marley10} 
fitted via MCMC.

\citet{Thurston00} and \citet{Marley10} considered Negative Binomial
additive models of the form:
\begin{equation}
\begin{array}{rcl}
\texttt{adducts}_{i}|\,\bbeta,\bu_1,\bu_2,\bu_3,\bu_4,\kappa &\simind&

\mbox{Negative-Binomial}(\exp\{ \beta_0 + f_1(\texttt{ageInit}_{i}) \\[2ex]
&& + f_2(\texttt{yearsSmoking}_{i}) + f_3(\texttt{yearsSinceQuit}_{i}) \\[2ex]
&& + f_4(\texttt{cigsPerDay}_{i}) \},\kappa),
\label{eq:adducts}
\end{array}
\end{equation}
with $\texttt{ageInit}_{i}$ the age of smoking initiation, 
$\texttt{yearsSmoking}_{i}$ the number of years of smoking, 
$\texttt{yearsSinceQuit}_{i}$ the number of years since quitting 
and $\texttt{cigsPerDay}_{i}$ the number of cigarettes smoked per day 
for subject $i$. The $f_{\ell}$, $1 \leq \ell \leq 4$, are modelled
using mixed-model based penalized splines as in (\ref{eq:OSull}),
with 20 basis functions each.

Figure \ref{fig:nbAdduct} displays the fitted functions for model
(\ref{eq:adducts}). \citet{Marley10} reported slow MCMC convergence
for this model, so we used burn-in size of 1000000
a retained sample size of 500000, and a thinning factor of 50
The MCMC-based fits are added as a reference to Figure \ref{fig:nbAdduct}.

Fitting of (\ref{eq:adducts}) via Algorithm \ref{alg:nbpBatchAlgorithmMFVB} 
took 2 minutes whilst MCMC fitting in \textsf{BUGS} took 1 hour and 28 minutes. 
As indicated by Figure \ref{fig:nbAdduct}, the much faster MFVB estimates
are quite close to the more accurate MCMC estimates. 

\ifthenelse{\boolean{ShowFigures}}
{
\begin{figure}[!ht]
\centering
{\includegraphics[width=\textwidth]{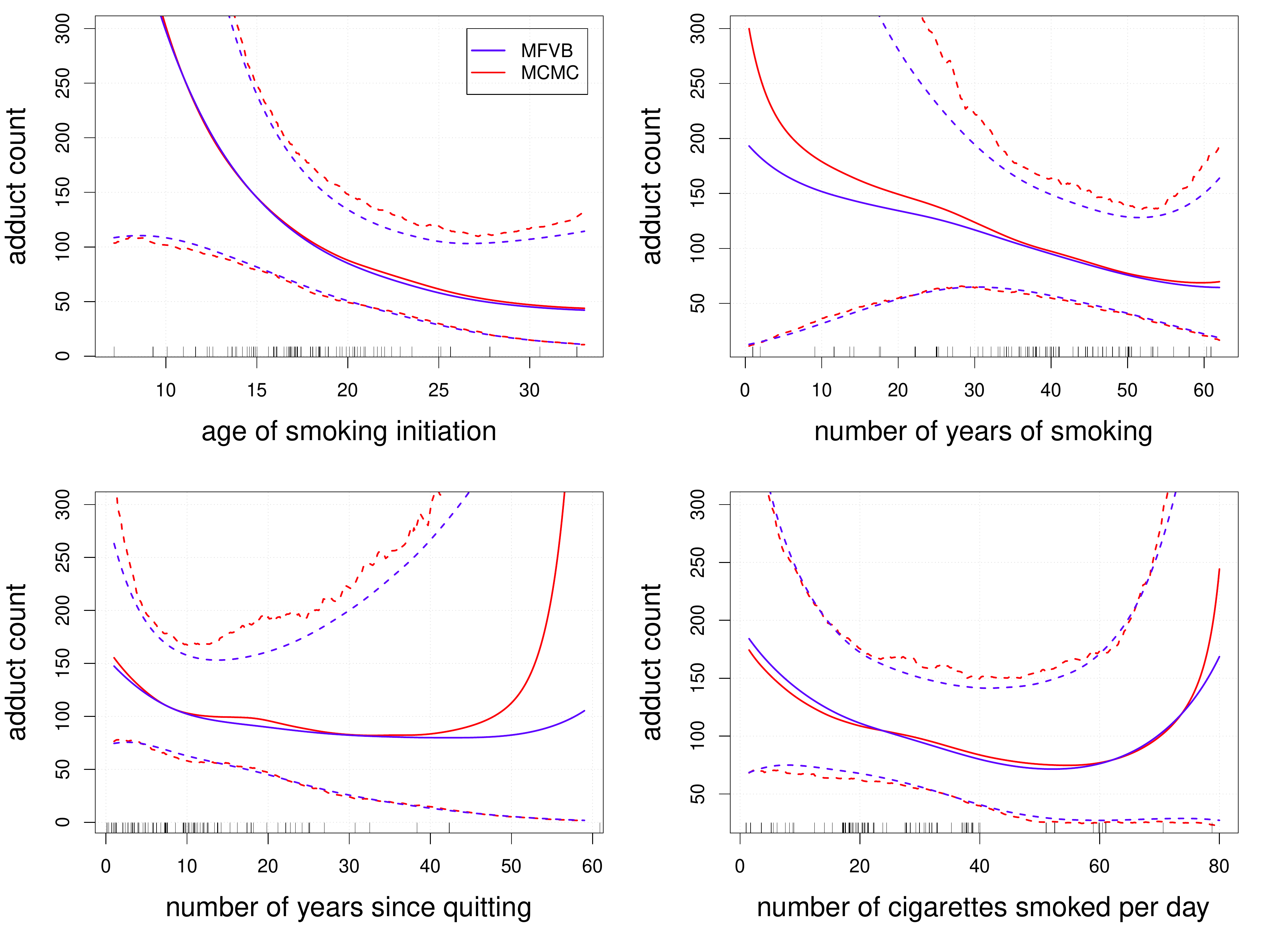}}
\caption{\it Negative Binomial regression result using MFVB and MCMC
  inference for adduct data set. Solid curves are posterior means for
  fitted functions while dashed curves are corresponding pointwise
  95\% credible sets.}
\label{fig:nbAdduct} 
\end{figure}
}
{\vskip3mm
\thickboxit{\bf \centerline{Negative Binomial result for GDELT data figure here.}}
\vskip3mm
}

\subsection{Real-time Poisson Nonparametric Regression Movie}\label{sec:movie}

The web-site \texttt{realtime-semiparametric-regression.net}
contains a movie that illustrates Algorithm \ref{alg:nbpOnlineAlgorithmMFVB}
in the special case of Poisson nonparametric regression with $r=1$
The spline basis functions set-up is analogous to that given in (\ref{eq:OSull}).

The data are simulated according to 
$$\xnew\sim\mbox{Uniform}(0,1),\quad
\ynew|\xnew\sim \mbox{Poisson}[\exp\{\cos(4\pi\xnew)+2\,\xnew\}]
$$
and the warm-up sample size is $\nwarm=100$.
The movie is under the link titled \texttt{Poisson nonparametric regression}.
and shows the efficacy of Algorithm \ref{alg:nbpOnlineAlgorithmMFVB}
for recovery of the underlying mean function in real time.

\section*{Appendix: Derivation of $q^*$ density functions}

\subsection*{Derivation of $q^*(a_{\ell})$ and $q^*(\sigma^2_{\ell})$ 
for the Poisson and Negative Binomial response model}

Standard manipulations lead to the following full conditional distributions:
$$
\begin{array}{l}
a_{\ell}|\text{rest} \simind
\mbox{Inverse-Gamma}(1,\sigma_{\ell}^{-2}+A_{\ell}^{-2})
\ \ \text{and} \\ [3ex]
\sigma^2_{\ell}|\text{rest} \simind 
\mbox{Inverse-Gamma}(1/2\,(K_{\ell}+1),a^{-1}_{\ell}+ 1/2 \, \Vert\bu_{\ell}\Vert^2).
\end{array}
$$

\subsection*{Derivation of the $(\bmu_{q(\bbeta,\bu)},\bSigma_{q(\bbeta,\bu)})$
 updates for the Poisson response model}

Adaptation of the derivations in Appendix A.3 of \citet{Wand13} leads to 
\begin{eqnarray*}
E_q\left[\log p(\by,\bbeta,\bu,\sigma^2_{1},\ldots,\sigma^2_{r},a_1,\ldots,a_r)\right]&=&
E_q\bigg[\log\,p(\by|\bbeta,\bu)+\log\,p(\bbeta,\bu|\sigma^2_{1},\ldots,\sigma^2_{r})\\
&& + \sum_{\ell=1}^{r}\log\,p(\sigma_\ell^2|a_\ell)+\sum_{\ell=1}^{r}\log\,p(a_\ell)\bigg]\\
&=&S+\mbox{terms not involving 
$\bmu_{q(\bbeta,\bu)}$ or $\bSigma_{q(\bbeta,\bu)}$}
\end{eqnarray*}
where
\begin{eqnarray*}
S&\equiv&
\by^T\bC\bmu_{q(\bbeta,\bu)}-\bone^T\exp\left\{\bC\bmu_{q(\bbeta,\bu)}
+\smhalf\diagonal(\bC\bSigma_{q(\bbeta,\bu)}\bC^T)\right\}\\
&&
-\smhalf\tr\left(\mbox{blockdiag}(\sigma_{\beta}^{-2}\bI_p,\mu_{q(1/\sigma_1^2)}
\bI_{K_1},\ldots,\mu_{q(1/\sigma_r^2)}\bI_{K_r})\{\bmu_{q(\bbeta,\bu)}
\bmu_{q(\bbeta,\bu)}^T+\bSigma_{q(\bbeta,\bu)}
\}\right)
\\
&&
-\smhalf P \log(2\pi)-\smhalf\,p\log(\sigma_{\beta}^2)
-\smhalf\,\sum_{\ell=1}^{r}K_\ell\,E_q\{\log(\sigma_\ell^2)\}-\bone^T\log(\by!).
\end{eqnarray*}
Then,
\begin{eqnarray*}
d_{\bmu_{q(\bbeta,\bu)}}\,S&=&\Big(\left[\by-\exp\{\bC\bmu_{q(\bbeta,\bu)}
+\smhalf\mbox{diagonal}(\bC\bSigma_{q(\bbeta,\bu)}\bC^T)\}\right]^T\bC\\
&&-\bmu_{q(\bbeta,\bu)}^T
\mbox{blockdiag}(\sigma_{\beta}^{-2}\bI_p,\mu_{q(1/\sigma_1^2)}
\bI_{K_1},\ldots,\mu_{q(1/\sigma_r^2)}\bI_{K_r})\Big)
d\bmu_{q(\bbeta,\bu)}\\  
\end{eqnarray*}
and by Theorem 6, Chapter 5, of \citet{Magnus99},
\begin{eqnarray*}
\{\Diff_{\bmu_{q(\bbeta,\bu)}}\,S\}^T
&=&\bC^T\left[\,\by-\exp\{\bC\bmu_{q(\bbeta,\bu)}
+\smhalf\mbox{diagonal}(\bC\bSigma_{q(\bbeta,\bu)}\bC^T)\}\right]\\
&&-\mbox{blockdiag}(\sigma_{\beta}^{-2}\bI_p,\mu_{q(1/\sigma_1^2)}
\bI_{K_1},\ldots,\mu_{q(1/\sigma_r^2)}\bI_{K_r})
\bmu_{q(\bbeta,\bu)}.\\
\end{eqnarray*}
Next,
\begin{eqnarray*}
d_{\vecof(\bSigma_{q(\bbeta,\bu)})}\,S&=&-\smhalf\vecof\Big(\bC^T
\diag[\exp\{\bC\bmu_{q(\bbeta,\bu)}
+\smhalf\mbox{diagonal}(\bC\bSigma_{q(\bbeta,\bu)}\bC^T)\}]\bC\\
&&+\mbox{blockdiag}(\sigma_{\beta}^{-2}\bI_p,\mu_{q(1/\sigma_1^2)}\bI_{K_1},\ldots,
\mu_{q(1/\sigma_r^2)}\bI_{K_r})\Big)^T
d\,\vecof(\bSigma_{q(\bbeta,\bu)})
\end{eqnarray*}
and
\begin{eqnarray*}
\vecof^{-1}\left((\Diff_{\vecof(\bSigma_{q(\bbeta,\bu)})}\,S)^T\right)
&=&-\smhalf(\bC^T\diag[\,\exp\{\bC\bmu_{q(\bbeta,\bu)}
+\smhalf\mbox{diagonal}(\bC\bSigma_{q(\bbeta,\bu)}\bC^T)\}]\,\bC\\
&&+\mbox{blockdiag}(\sigma_{\beta}^{-2}\bI_p,\mu_{q(1/\sigma_1^2)}\bI_{K_1},\ldots,
\mu_{q(1/\sigma_r^2)}\bI_{K_r})).
\end{eqnarray*}
The final result follows from plugging in 
$\{\Diff_{\bmu_{q(\bbeta,\bu)}}\,S\}^T$ and 
$\vecof^{-1}\left((\Diff_{\vecof(\bSigma_{q(\bbeta,\bu)})}\,S)^T\right)$
in the updating formulas (\ref{eq:updatesMatt}).

\subsection*{Derivation of $q^*(g_i)$ and $q^*(\kappa)$ for the Negative Binomial response model}

Standard manipulations lead to the following full conditional distribution
$$
\begin{array}{l}
g_i|\text{rest} \simind \mbox{Gamma}(\kappa+y_i,1+\kappa \exp\{ -\bc_i^T [\bbeta^T \ \bu^T]^T\})
\end{array}
$$
such that $q^*(g_i)$ is the Gamma density function specified 
in (\ref{eq:nbOptimalQ}). In addition, standard distributional results
for the Gamma density function lead to 
\begin{equation}
\begin{array}{ll}
\bmu_{q(\log(\bg))} = &\text{digamma}(\bone\mu_{q(\kappa)}+\by) - 
\log( \bone + \mu_{q(\kappa)} \exp \{ -\bC \bmu_{q(\bbeta,\bu)} \\[2ex]
& + \frac{1}{2} \, \mbox{diagonal}( \bC \bSigma_{q(\bbeta,\bu)}
\bC^T) \} ). 
\nonumber
\end{array}
\end{equation}
The density function $q^*(\kappa)$ can be obtained by adapting the
expressions in Appendix A.1 of \citet{Wand11} and result in
$$
\mu_{q(\kappa)} = \exp\left[
  \log\left\{\Hsc(1,n,C_1,\kappa_{\text{min}},
\kappa_{\text{max}}) \right\} - 
\log\left\{\Hsc(0,n,C_1,\kappa_{\text{min}},\kappa_{\text{max}})
 \right\} \right].$$

\subsection*{Derivation of  the $(\bmu_{q(\bbeta,\bu)},\bSigma_{q(\bbeta,\bu)})$
 updates for the Negative Binomial response model}

Note that 
\begin{eqnarray*}
E_q[\log p(\by,\bg,\bbeta,\bu,\kappa,\sigma^2_{1},\ldots,\sigma^2_{r},a_1,\ldots,a_r)]&=&
E_q\bigg[\log\,p(\by|\bg)+\log\,p(\bg|\bbeta,\bu,\kappa)\\
&&+\log\,p(\bbeta,\bu|\sigma^2_{1},\ldots,\sigma^2_{r})+\log\,p(\kappa)\\
&& + \sum_{\ell=1}^{r}\log\,p(\sigma_\ell^2|a_\ell)+\sum_{\ell=1}^{r}\log\,p(a_\ell)\bigg]\\
&=&S+\mbox{terms not involving 
$\bmu_{q(\bbeta,\bu)}$ or $\bSigma_{q(\bbeta,\bu)}$}
\end{eqnarray*}
where
\begin{equation}
\begin{array}{lll}
S &\equiv& n E_q [\kappa \log \left(\kappa\right)] - \mu_{q(\kappa)}
\bone^T \bC \bmu_{q(\bbeta,\bu)} - n E_q [
\log\left(\Gamma(\kappa)\right) ] + (\mu_{q(\kappa)}-1) \bone^T 
E_q [ \log \left( \bg \right) ] \\[2ex]
&& - \mu_{q(\kappa)} \bmu^T_{q(\bg)} \exp\{-\bC\bmu_{q(\bbeta,\bu)}
+\smhalf\diagonal(\bC\bSigma_{q(\bbeta,\bu)}\bC^T)\}\\[2ex]
&&
-\smhalf\tr\left(\mbox{blockdiag}(\sigma_{\beta}^{-2}
\bI_p,\mu_{q(1/\sigma_1^2)}\bI_{K_1},\ldots,\mu_{q(1/\sigma_r^2)}
\bI_{K_r})\{\bmu_{q(\bbeta,\bu)}\bmu_{q(\bbeta,\bu)}^T+\bSigma_{q(\bbeta,\bu)}
\}\right)\\[2ex]
&&
-\smhalf P \log(2\pi)-\smhalf\,p\log(\sigma_{\beta}^2)
-\smhalf\,\sum_{\ell=1}^{r}K_\ell\,E_q\{\log(\sigma_\ell^2)\}. \nonumber
\end{array}
\end{equation}
Then,
\begin{eqnarray*}
\{\Diff_{\bmu_{q(\bbeta,\bu)}}\,S\}^T
&=& \mu_{q(\kappa)} \bC^T\left[\,\bmu_{q(\bg)} \odot \exp\{-\bC\bmu_{q(\bbeta,\bu)}
+\smhalf\mbox{diagonal}(\bC\bSigma_{q(\bbeta,\bu)}\bC^T)\}-\bone\right]\\
&&-\mbox{blockdiag}(\sigma_{\beta}^{-2}\bI_p,\mu_{q(1/\sigma_1^2)}
\bI_{K_1},\ldots,\mu_{q(1/\sigma_r^2)}\bI_{K_r})
\bmu_{q(\bbeta,\bu)}\\
\end{eqnarray*}
and 
\begin{eqnarray*}
d_{\vecof(\bSigma_{q(\bbeta,\bu)})}\,S
&=&-\smhalf\vecof\Big(\mu_{q(\kappa)} \bC^T\diag[\bmu_{q(\bg)} \odot 
\exp\{-\bC\bmu_{q(\bbeta,\bu)}\\
&&+\smhalf\mbox{diagonal}(\bC\bSigma_{q(\bbeta,\bu)}\bC^T)\}]\bC\\
&&+\mbox{blockdiag}(\sigma_{\beta}^{-2}\bI_p,\mu_{q(1/\sigma_1^2)}\bI_{K_1},
\ldots,\mu_{q(1/\sigma_r^2)}\bI_{K_r})\Big)^T
d\,\vecof(\bSigma_{q(\bbeta,\bu)})
\end{eqnarray*}
such that
\begin{eqnarray*}
\vecof^{-1}\left((\Diff_{\vecof(\bSigma_{q(\bbeta,\bu)})}\,S)^T\right)
&=&-\smhalf(\mu_{q(\kappa)} \bC^T\diag[\bmu_{q(\bg)} \odot \exp\{-\bC\bmu_{q(\bbeta,\bu)}\\
&&+\smhalf\mbox{diagonal}(\bC\bSigma_{q(\bbeta,\bu)}\bC^T)\}]\,\bC\\
&&+\mbox{blockdiag}(\sigma_{\beta}^{-2}\bI_p,\mu_{q(1/\sigma_1^2)}\bI_{K_1},
\ldots,\mu_{q(1/\sigma_r^2)}\bI_{K_r})).
\end{eqnarray*}
The final result follows from plugging in these expressions in the
updating formulas (\ref{eq:updatesMatt}).

\section*{Acknowledgments}

This research was partially supported by Australian Research Council 
Discovery Project DP110100061. The authors are grateful to Marianne Menictas
for her comments on this research.

\setstretch{1.0}

\end{document}